\documentclass[12pt]{article}
\usepackage{ssi}
\usepackage{times}
\usepackage{epsf}
\usepackage{color}

\newcommand{\cbla}{\color{black}}
\newcommand{\cred}{\color{red}}

\newcommand{\ltsim}{\raisebox{-1ex}
{\mbox{$\mbox{}\stackrel{\textstyle <}{\sim}\mbox{}$}}}

\begin{document}

\begin{flushright}
Fermilab-Conf-00-347-E \\
\end{flushright}

\title{Electroweak and $B$ Physics Results from the Fermilab Tevatron 
Collider}

\author{Kevin T. Pitts\thanks{Work supported by the Department of
Energy, Contract DE-FG02-91ER40677.
\vskip 0.5in 
\noindent
 }\\ 
 University of Illinois, Department of Physics \\
1110 West Green Street, Urbana, IL 61801-3080, USA \\
{\it E-mail: kpitts@uiuc.edu}\\[0.4cm]
Representing the CDF and D\O\  Collaborations
}

\maketitle
\begin{abstract}
\baselineskip 16pt 
This writeup is an introduction to 
some of the experimental issues
involved in performing electroweak
and $b$ physics measurements at the Fermilab
Tevatron.   In the electroweak sector, we
discuss $W$ and $Z$ boson
cross section measurements as well as the
measurement of the mass of the $W$ boson.
For $b$ physics, we discuss measurements of
$B^0/\overline{B^0}$ mixing and $CP$ violation.
This paper is geared towards 
nonexperts
who are interested in understanding some of the
issues and motivations for these measurements 
and how the measurements are carried out.

\end{abstract}

\section{Introduction}

The Fermilab Tevatron collider is currently  between data runs.
The period from 1992-1996, known as Tevatron Run~1, saw
both the CDF and D\O \ experiments accumulate approximately
$110\, \rm pb^{-1}$ of integrated luminosity.  These
data sets have yielded a large number of results and
publications on topics ranging from the discovery of the
top quark to precise measurements of the mass of the $W$
boson;  from  measurements of jet production at the highest 
energies ever observed to searches for physics beyond 
the Standard Model.  

 This talk and subsequent  paper focus on two aspects
of the Tevatron program: electroweak physics and the 
physics
of hadrons containing the bottom quark.  Each of these
topics is quite rich in its own right.  It is  not
possible to 
do justice to either these topics in the
space provided.   

Also, there are a large number of
sources for summaries of recent results.  For example,
many conference proceedings and summaries are easily
accessible to determine the most up-to-date measurements
of the mass of the $W$ boson.   Instead of trying to
summarize a boat-load of Tevatron measurements here, I 
will attempt to describe a few  measurements in an
introductory manner.
The goal of this paper is to explain some of
the methods and considerations  for these
measurements. 
This paper therefore is 
geared more towards
students and non-experts. 
The goal here is not to comprehensively 
present the results, but to 
discuss  how the
results are obtained and what the important elements
are in these measurements.

After a brief discussion of the Tevatron collider
and the two collider experiments, we will discuss
electroweak and $b$ physics at the Tevatron.

\section{The Tevatron Collider}

The Fermilab Tevatron collides protons($p$) and antiprotons
($\overline{p}$) at very high energy.   In past runs,
the $p\overline{p}$ 
center of mass energy was $\sqrt{s} = 1.8\, \rm TeV$.
It will be increased in the future to $2\, \rm 
TeV$.\footnote{For the upcoming Tevatron run, the 
center of mass energy will be $\sqrt{s} = 1.96\, \rm TeV$.
Running the machine at slightly below $2\, \rm TeV$ 
drastically improves the reliability of the superconducting
magnets.}
Until the Large Hadron Collider begins operation at
CERN late in this decade, 
the Tevatron will be the highest energy 
accelerator in the world.  The high energy, combined
with a very high interaction rate, provides many
opportunities for unique and interesting measurements.

\begin{table}[thb]
\begin{center}
\caption{Some highlights in the history of the 
Fermilab Tevatron.  This table lists primarily
milestones associated with the collider program.
In addition, there have been several  Tevatron
fixed-target runs, producing a wealth of
physics results.}
\label{ta:tev} 
\begin{tabular}{ll}
\hline
 1969 &  
ground breaking for
 National Accelerator Laboratory ``Main Ring'' \\
 1972 &  
200 GeV beam in the Main Ring \\
 1983 &  
first beam in the ``Energy Doubler'' $\Rightarrow$ ``Tevatron''  \\ 
 1985 &
CDF observes first $p\overline{p}$ collisions  \\
 1988-89 &  
Run 0, CDF collects $\sim $3$\, \rm pb^{-1}$ \\
 1992-93 &  
Run 1A, CDF and D\O \ collect $\sim $20$\, \rm pb^{-1}$ \\ 
 1994-95\ \   &  
Run 1B, CDF and D\O \ collect $\sim $90$\, \rm pb^{-1}$ \\
 2001-02 &  
 Run 2 with new Main Injector and Recycler, \\
 & upgraded CDF and D\O\  expect 2000$\, \rm pb^{-1}$=2$\rm \, fb^{-1}$ \\
 2003- &   Run 3, 15-30$\, \rm fb^{-1}$ \\
\hline
\end{tabular}
\end{center}
\end{table}

The Tevatron has a history that goes back over $20$ years.
Table~\ref{ta:tev} lists a few of the highlights.  The
original Fermilab accelerator, the ``Main Ring'', was finally
decommissioned in 1998 after more than 25 years of operation.
In collider mode, the Main Ring served as an injector for
the Tevatron.  The Main Ring and Tevatron resided in the same
tunnel of  circumference of  $\sim \! 4\,$miles.  The Tevatron now
resides alone in this tunnel.

The Tevatron  consists of approximately
 1000 superconducting magnets.
Dipole magnets  are $\sim \! 7\, \rm m$ in length, cooled
by liquid helium to 
 a temperature of $3.6\, \rm K$ and typically carry currents
of over $4000\, \rm Amps$.
Protons and anti-protons are injected into the Tevatron at
an energy of $150\, \rm GeV$, then their energy is raised to
the nominal energy which
was $900\, \rm GeV$ per beam in the past and will be 
$980\, \rm GeV$ per beam for the upcoming run.
During the period known as Run 1B, the
Tevatron routinely achieved a luminosity that was 
more than 20 times the original design luminosity 
of $10^{30}\, \rm cm^{-2} s^{-1}$ \cite{tev}.

The major upgrade in recent years has been
the construction of the Main Injector which replaces
the Main Ring.  The Main Injector, along
with another new accelerator component, the Recycler, will
allow for much higher proton and antiproton intensities,
and therefore higher luminosity than previously achievable.
The anticipated Tevatron luminosity in the upcoming run
will be a factor of $200$ beyond the original design
luminosity for the Tevatron.

The CDF and D\O\ results presented here are from the 
$110\, \rm pb^{-1}$ of integrated luminosity collected
in the period of $1992$-$1996$.  The expectations for 
Run~II are 
for a  $20$-fold increase in the data sample by $2003$
($2\, \rm fb^{-1}$).  Beyond Run~II, the goal is to increase
the data sample by an additional factor of $10$
($15$-$30\, \rm fb^{-1}$)
by the time that the LHC begins producing results.

\section{CDF and D\O}

The CDF and D\O\ detectors are both axially symmetric
detectors that cover about $98\% $ of the full $4\pi$
solid angle around the proton-antiproton interaction
point.
The experiments utilize similar strategies for 
measuring the interactions.  Near the interaction
region, tracking systems accurately measure the 
trajectory of charged particles.  Outside the 
tracking region, calorimeters surround the interaction
region to measure the energy of both the charged and
neutral particles.  Behind the calorimeters are 
muon detectors, that measure the deeply penetrating
muons.
Both experiments have fast trigger and readout
electronics to acquire data at high rates.
Additional details about the experiments can be found
elsewhere \cite{cdf,d0}.  

The strengths of the detectors are somewhat 
complementary to one another.
The D\O\ detector features a uranium liquid-argon
calorimeter that has very good energy resolution
for electron, photon and hadronic jet energy measurements.
The CDF detector features a $1.4\, \rm T$ solenoid
surrounding a silicon microvertex detector and
gas-wire drift chamber.  These properties, combined
with muon detectors and calorimeters, allow
for excellent muon and electron identification, as
well as precise tracking and vertex detection
for $B$ physics.

\section{Electroweak Results}

Although many precise electroweak measurements have
been performed at and above the $Z^0$ resonance at
LEP and SLC, the Tevatron provides some unique and
complementary measurements of electroweak phenomena.
Some of these measurements include $W$ and $Z$ 
production cross sections; gauge boson couplings 
($WW$,$W\gamma$,$WZ$,$Z\gamma$,$ZZ$); and properties
of the $W$ boson  (mass, width, asymmetries).

For the most part, both $W$ and $Z$ bosons are observed
in hadron collisions through leptonic decays to electrons
and muons, such as  $W^+\rightarrow e^+\bar{\nu}_e$ and
$Z^0\rightarrow \mu^+\mu^-$.
 The branching ratios for
the leptonic decays of the $W$ and $Z$ are significantly
smaller than the branching ratios for hadronic decays.  There are about
3.2 hadronic $W$ decays for every $W$ decay to $e$ or $\mu$
and  about 10 hadronic $Z$ decays for every
$Z$ decay to $e^+e^-$ or $\mu^+\mu^-$.  Unfortunately, 
the dijet background from processes like
$qg\rightarrow qg$  and 
$gg\rightarrow q\overline{q}/gg$ (in addition to higher order processes)
totally swamp the signal from $Z^0\rightarrow q\overline{q}$
and $W^+\rightarrow q\overline{q}^\prime$.\footnote{There are 
special cases where hadronic decays of heavy gauge
bosons have been observed:  hadronic 
$W$ boson decays have been 
observed in top quark decays, and a $Z^0\rightarrow b\overline{b}$
signal has been observed by CDF.  Also, both experiments 
have observed $W$ and $Z$ decays to $\tau$ leptons.}

\subsection{$W$ and $Z$ Production}

The rate of production of $W$ and $Z$ bosons is an
interesting test of the theories 
of both electroweak and strong interactions.  
The actual production rates are determined by factors that
include the gauge boson couplings to fermions (EW) and
the parton distribution functions and higher order
corrections (QCD).

As an
example analysis, we will discuss the  measurement 
the $Z$ production cross-section
from the $Z^0\rightarrow e^+e^-$ mode.  The total
number of events we observe will be:
\begin{equation}
N = {\cal L}_{int}\cdot \sigma_Z \cdot 
Br(Z^0\rightarrow e^+e^-)\cdot \epsilon_{ee}
\label{eq:1}
\end{equation}
where ${\cal L}$ is the instantaneous luminosity,
${\cal L}_{int} = \int {\cal L}dt$ is the integrated luminosity,
$\sigma_Z = \sigma(p\overline{p}\rightarrow Z^0X)$ is
the $Z$ boson production cross section, 
$Br(Z^0\rightarrow e^+e^-)$ is the branching ratio
for $Z^0\rightarrow e^+e^-$,
and 
$\epsilon_{ee}$ is the efficiency for observing
this decay mode.   We have made the simplifying 
assumption that there are  no background events
in our signal sample.   Let's take each term in turn:
\begin{itemize}
\item ${\cal L}_{int} = \int {\cal L}dt $: 
the integrated luminosity is measured
in units of $\rm cm^{-2}$ and is a measure of the
total number of $p\overline{p}$ interactions.  The 
instantaneous luminosity is measured in units
of $\rm cm^{-2}s^{-1}$.  In this 
case, ``integrated'' refers to the total time 
the detector was ready and able to measure
$p\overline{p}$ interactions.\footnote{We refer to the
detector as ``live'' when it is ready and available
to record data.   If the detector is off or busy processing
another event, it is not available or able to record
additional data.  This is known as ``dead-time''.}

\item $\sigma_Z = \sigma(p\overline{p}\rightarrow Z^0X)$:
cross sections are measured in units of $\rm cm^{2}$ and
are often quoted in units of ``barns'', where $\rm 1 b = 10^{-24} \rm
cm^{2}$.  Typical electroweak 
 cross sections measured at the Tevatron are
in nanobarns ($\rm nb = 10^{-9}\rm b$)  or
picobarns ($\rm pb = 10^{-12} b$).  The \bf total \rm cross
section for $p\overline{p}$ at the Tevatron is about $70\, \rm mb =
70\times 10^{-3}\rm b$.  The cross section listed here is for
any and all types of $Z$ boson production.  The ``$X$'' 
includes the remaining fragments of the initial $p$ and
$\overline{p}$, in addition to allowing for additional final
state particles.
\item $Br(Z^0\rightarrow e^+e^-)$:  The branching ratio
is the fraction of $Z^0$ bosons that decay to a specific
final state, $e^+e^-$ in this example.\footnote{The branching
ratio is the fraction of times 
that  a particle will 
decay into a specific final state.  More concisely,
the branching ratio is $Br(Z^0\rightarrow e^+e^-) =
\Gamma(Z^0\rightarrow e^+e^-)/\Gamma(Z^0\rightarrow
\rm all)$, where $\Gamma(Z^0\rightarrow e^+e^-)$ is
the partial width for $Z^0$ decaying to $e^+e^-$
and $\Gamma(Z^0 \rightarrow \rm all)$ is the total
$Z^0$ width.}  
\item $\epsilon_{ee}$:  Of the $Z^0$ bosons that are produced
and decay to $e^+e^-$, not all of them are detected or accepted
into the final event sample.  Some of the events are beyond the
region of space the detector covers in addition to the fact
that the detector is not 100\% efficient for detecting any
signature.
\end{itemize}

Our ultimate goal is to extract  $\sigma_Z$.  Rearranging
Equation~\ref{eq:1}, we have:
\begin{equation}
\sigma_Z = {N\over{{\cal L}_{int}\cdot 
Br(Z^0\rightarrow e^+e^-)\cdot \epsilon_{ee}}}.
\end{equation}
From the data, we can count the number of signal events, $N$.
To extract a cross section, we need to know the terms in the
denominator as well:
\begin{itemize}
\item The luminosity is measured by looking at the total
rate for $p\overline{p}\rightarrow p\overline{p}X$ in a 
specific and well-defined detector region.  This
rate is measured as a function of time and then integrated
over the time the detector is live.  The equation 
$N = {\cal L} \sigma$ is used again, in this case we already
know the total $p\overline{p}$ cross section($\sigma$), 
so we can use this
equation to extract ${\cal L}$.   At $e^+e^-$ machines, 
the measurement of the luminosity is quite precise, with
a relative  error of $1\%$ or less.  For hadron machines, 
that level of precision is  not possible.  Typical relative
uncertainties on the luminosity are $5$-$8\%$\cite{derwent}.
\item The branching ratio for $Z^0\rightarrow e^+e^-$ is
measured quite precisely by the LEP and SLC experiments.
The world average value is used as an input here.  The
uncertainty on that value is incorporated into the ultimate
uncertainty on the cross section.
\item The efficiency for a final state like this is measured
by a combination of simulation and control data samples.
Primarily, data samples are used that are well understood.
For example, $Z^0$ decays ($Z^0\rightarrow e^+e^-$ and 
$Z^0\rightarrow \mu^+\mu^-$)
provide an excellent sample of electrons and muons 
for detector calibration.
The high invariant mass of the lepton 
pair is a powerful handle to
reject background.
\end{itemize}

\begin{figure}[ht]
\begin{center}
\epsfxsize=3.5in 
\hspace*{0.4truein}
\epsfbox{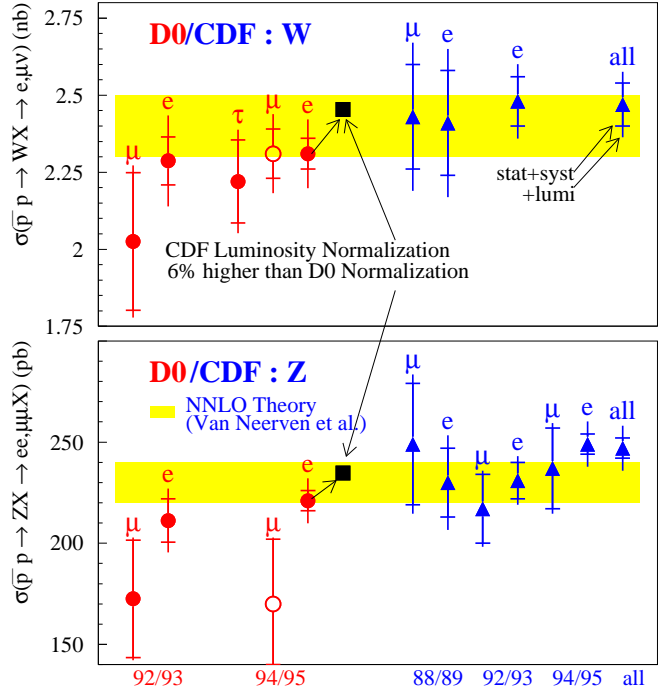}
\caption{Summary of D\O\ and CDF $W$ and $Z$ boson cross 
section measurements.  The solid bands indicate the theoretical
prediction.  The circular points are the D\O\ results; the 
triangles are the CDF results.  The two experiments use a
different luminosity normalization.
}
\label{fig:wzxsec}
\end{center}
\end{figure}

Putting all of these factors together, it is possible to
measure the total cross sections for $p\overline{p}\rightarrow
WX$ and $p\overline{p}\rightarrow ZX$.   These measurements
are  performed independently in both electron and muon modes.  
However, after the corrections 
for the efficiencies of each mode, the measurements should
(and do) yield consistent  measured values for the production
cross section.

The results from D\O\ and CDF are represented in 
Fig.~\ref{fig:wzxsec}.   The top plot is for $W$
production, the bottom plot for $Z$ production.  The
shaded region is the theoretical cross section.  On
both plots, the circular points are the D\O\ measurements,
the triangles the CDF measurements.   Part of the difference
in the results from the two experiments arises from a 
different calculation of ${\cal L}_{int}$.  If a 
common calculation were used, the D\O\ numbers would be
$6\%$ larger than those presented.  This shows that in
fact the integrated luminosity is the largest systematic
uncertainty on the cross sections.  Details of these 
analyses may be found in the literature~\cite{cdfxsec,d0xsec}.

\subsection{${\cal R}$ and the $W$ Width}

One way to make the measurement more sensitive to
the electroweak aspects of the $W$ and $Z$ production
processes is to measure the
cross section ratio.  This ratio is often referred to as ``${\cal R}$'',
and defined as:
\begin{displaymath}
{\cal R} \equiv \frac{\sigma(W)}{\sigma(Z)} \cdot \frac{Br(W\rightarrow
\ell \nu)}{Br(Z\rightarrow \ell \ell )}
\end{displaymath}

In taking the ratio of cross sections, the integrated
luminosity  (${\cal L}_{int}$) term
and its uncertainty cancel.  Other experimental and 
theoretical uncertainties cancel as
well, making the measurement of ${\cal R}$ a more stringent
test of the Standard Model.  As we can see from Fig.~\ref{fig:wzxsec},
the ratio is about equal to $10$.
This is confirmed by the results shown in Table~\ref{ta:r}.
\begin{table}[ht]
\begin{center}
\caption{Summary of Tevatron measurements of ${\cal R}$,
where ${\cal R} \equiv \frac{\sigma(W)}{\sigma(Z)} \cdot 
\frac{Br(W\rightarrow
\ell \nu)}{Br(Z\rightarrow \ell \ell )}$.  
}
\label{ta:r} 
\begin{tabular}{ll}
\hline
& measured value of ${\cal R}$ \\
\hline
\rm D\O\  &  $10.43\pm 0.15 
(\rm stat.) \pm 0.20 (\rm syst.) 
\pm 0.10 (\rm theory)$ \\
\rm CDF & $10.38 \pm 0.14 (\rm stat.) 
\pm 0.17 (\rm syst.)$ \\
\hline
\end{tabular}
\end{center}
\end{table}
The D\O\ result is for the electron final state \cite{d0r}; 
the CDF result is
for the electron and muon final states \cite{cdfr}.  For the CDF result,
the theoretical uncertainty is contained in the systematic uncertainty.

\begin{figure}[t!]
\begin{center}
\epsfxsize=5.5in 
\epsfbox{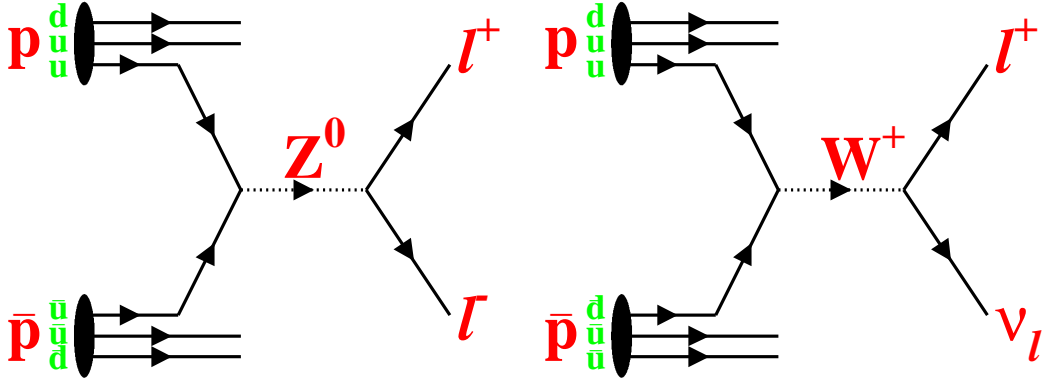}
\caption{Example $Z$ and $W$ production diagrams in
high energy $p\overline{p}$ collisions.  These figures
show valence-valence production, which in fact is  not
dominant at Tevatron energies.  The dominant production
mechanism is quark-antiquark annihilation, where one
quark(antiquark) is a valence quark and the other 
antiquark(quark) is a 
sea quark.
}
\label{fig:wprod}
\end{center}
\end{figure}

We can take this result one step further.   The measured 
quantity is ${\cal R}$.  Theoretically, the cross section
ratio $\sigma(W)/\sigma(Z)$ is calculated with good
precision.  This can be understood by noting that the 
primary production of $Z$ bosons at the Tevatron arise
from the reactions:
$u\overline{u}\rightarrow Z^0$   and   
$d\overline{d}\rightarrow Z^0$,
where the up and down quarks (and antiquarks) can be valence 
or sea quarks in the proton.  An example of valence-valence
production is shown in Fig.~\ref{fig:wprod}.
For $W$ production, the primary contributions are
 $u\overline{d} \rightarrow W^+$ and 
 $\overline{u}d \rightarrow W^-$.
These reactions look quite similar to the $Z$ production
mechanisms where a $u$ quark is replaced with a $d$
quark (or vice-versa).  An example of valence-valence
$W^+$ production is also shown in Fig.~\ref{fig:wprod}.

Although 
both $Z^0$ and $W^\pm$ are produced through
quark-antiquark annihilation, the dominant contribution
is not from the valence-valence diagrams shown in 
Fig.~\ref{fig:wprod}.  The typical $q\overline{q}$ 
interaction energy for heavy boson production is the
mass of the boson: $\sqrt{\hat{s}}\sim M_{Z,W}$.
Since the heavy boson mass $M_{Z,W}\sim \! 100\, \rm GeV =
0.1 \, \rm TeV$ and the $p\overline{p}$ center of mass
energy is $\sqrt{s}\sim \! 2\, \rm TeV$, the
process 
requires the $q\overline{q}$ center of mass energy to be
only $\sqrt{\hat{s}}/\sqrt{s}\simeq 5\%$ of the 
$p\overline{p}$ center of mass energy.
In other words, if a quark and antiquark are each carrying
$5\%$ of the proton (and antiproton) momentum, then there
is sufficient collision energy to produce a heavy boson.

Both valence and sea quarks have a good probability for
carrying a sufficient fraction of the proton's energy to
produce a gauge boson.   In fact, the dominant production
mechanism at the Tevatron 
is annihilation where the quark(antiquark) is a valence 
quark and the antiquark(quark) is a sea quark.  
The valence-sea
production mechanism 
is about $4$ times larger than the valence-valence
and sea-sea production mechanisms.  It is coincidental
that the valence-valence and sea-sea mechanisms are about
equal at this energy.  At higher energies, the sea-sea 
mechanism dominates; at lower energies, the valence-valence
mechanism dominates \cite{scott}.

The theoretical predictions for 
the  production cross sections of  $Z$ and
$W$ bosons are not known to high precision.  Strong
interaction effects, such as the 
parton distribution
functions and higher order diagrams lead to
theoretical uncertainty.  
The ratio of cross sections is well
calculated, however,  because going from $Z$ production to 
$W^+$ production amounts to replacing an $\overline{u}$
with a $\overline{d}$.  
In addition,
the gauge boson couplings to fermions are well 
measured.   Combining these points makes the 
ratio of cross sections a much better determined
quantity than the individual cross sections.

\begin{figure}[t]
\begin{center}
\epsfxsize=3in 
\hspace*{0.4truein}
\epsfbox{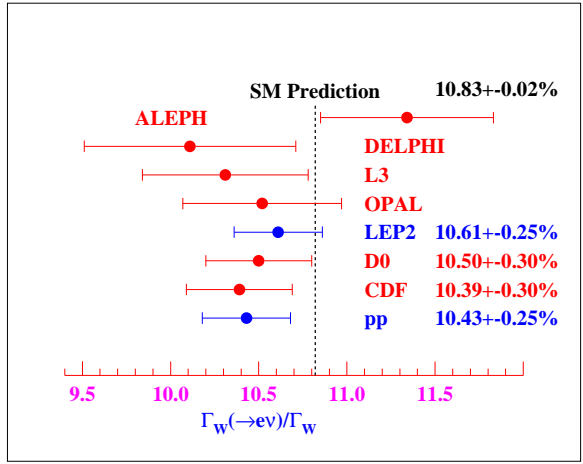}
\caption{Measurements of the branching ratio for
$W\rightarrow e\nu_e$.  The Tevatron results come
from a measurement of ${\cal R}$ combined with the 
LEP measurement of $Br(Z^0\rightarrow e^+e^-)$ and a
theoretical calculation of $\sigma(W)/\sigma(Z)$.
}
\label{fig:brw}
\end{center}
\end{figure}

Additionally, the branching ratio for $Z^0 \rightarrow
\ell^+ \ell^-$ is well measured at LEP.  Using
our measured value of ${\cal R}$, inputting the
theoretical value for $\sigma(W)/\sigma(Z)$ and using
the LEP value for $Br(Z^0\rightarrow \ell^+\ell^-)$,
we can extract the branching ratio for $W\rightarrow \ell \nu$.
This is shown in Fig.~\ref{fig:brw}.
The Tevatron results have similar uncertainties to the 
results from LEP2.  As the uncertainties are reduced,
this measurement will continue to be an important test 
of the Standard Model.

\subsection{$W$ mass}

The electroweak couplings and boson masses within the
Standard Model may be completely specified by three 
parameters.
Typically, those parameters are chosen to be
$M_Z$ (the mass of the $Z^0$ boson),
$G_F$, (the Fermi constant), and 
$\alpha_{QED}$ (the electromagnetic
coupling constant).
These
three parameters are not required to be the inputs,
though.  For example, we could choose to use the
charge of the electron ($e$), the weak mixing angle
($\sin^2\theta_W$) and the mass of the $W$ boson 
($M_W$) as our inputs.  At tree level (no radiative
corrections, also known as Born level), 
any set of three parameters is sufficient to calculate
the remaining quantities.  The three chosen:
$M_Z$, $G_F$ and $\alpha_{QED}$ are the ones measured
experimentally with the highest precision.

\begin{figure}[ht]
\begin{center}
\epsfxsize=4.75truein 
\hspace*{0.2truein}
\epsfbox{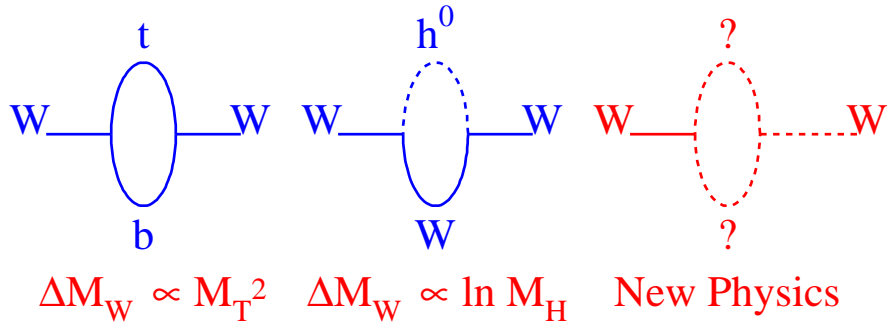}
\caption{Loop contributions to the $W$ mass.  The $\Delta M_W$
denotes the shift in $W$ mass from the Born level value.  The
dependence upon the top quark mass is more dramatic than the
dependence on the Higgs mass.  New physics can appear in these
loop corrections as well.}
\label{fig:mwrad}
\end{center}
\end{figure}

Therefore, at Born level, these three parameters are 
sufficient to 
exactly determine the mass of the $W$ boson.  The true
$W$ mass depends additionally on radiative corrections,
the most important of which involve the top quark and
the Higgs boson.  Radiative corrections involving fermion
or boson
loops grow with the mass of the particle in the loop.  This
is why the top quark and Higgs boson masses are the
most important corrections to the $W$ mass.  
These loop diagrams  are  shown graphically in 
Fig.~\ref{fig:mwrad}.

The $W$ mass can be calculated with a high degree
of precision and therefore simply measuring the
$W$ mass provides a test of the Standard Model.  
Since there is additional uncertainty on the $W$
mass due to the unknown mass of the Higgs boson
(or perhaps it doesn't exist in the Standard 
Model form) the simple test of comparing the 
measured $W$ mass value to the prediction is not
a high precision test.
It is an important test, though, because
deviations from the Standard Model 
predicted $W$ mass can
arise through other non-Standard Model 
particles affecting the $W$
mass through loops.  

\begin{figure}[t]
\begin{center}
\epsfxsize=3.75truein 
\hspace*{0.2truein}
\epsfbox{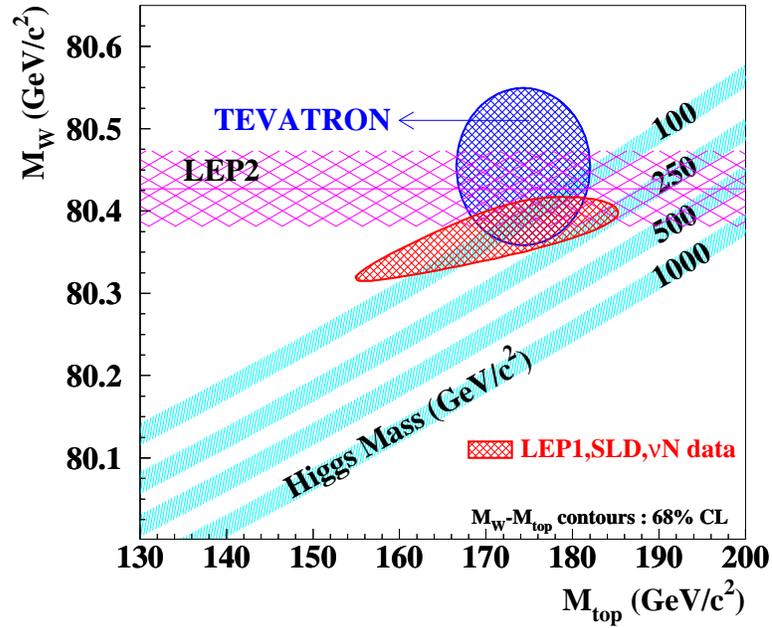}
\caption{
A summary of precision electroweak measurements.  The Tevatron
contour is from direct measurements of $M_W$ and $M_t$.  The
horizontal band is the direct measurement of $M_W$ from LEP2.  The oval
contour arises from  precision electroweak measurements of 
$\sin^2\theta_W^{eff}$, and $Z$ couplings and asymmetries
translated into the $M_W$, $M_t$ plane.  The bands are Standard
Model contours for various values of the Higgs mass, $M_H$.  It
is clear that the experimental results are consistent
with one another and currently favor a light
Higgs boson.} 
\label{fig:mwmt}
\end{center}
\end{figure}

In addition, when combined with the measured value
for the top quark mass ($M_t$), we can constrain the
Higgs mass.  In saying that we can constrain the Higgs
mass, this is implicitly assuming a Standard Model
Higgs boson.    This can be seen graphically in
Fig~\ref{fig:mwmt}, where electroweak results are
plotted in the $M_W$,$M_t$ plane.  The contour
marked ``Tevatron'' shows the directly measured values
for $M_W$ and $M_t$.  The bands are contours of Standard
Model calculations for $M_W$ versus $M_t$ for different
masses of the Higgs boson.  The current Tevatron 
region is consistent with the Standard Model and prefers
a light Higgs boson.

Another way the $W$ mass tests
the Standard Model is through self-consistency with
other Standard Model measurements.  For example, 
the LEP1, SLD, $\nu$N data contour in Fig.~\ref{fig:mwmt}
arises from taking the electroweak measurements of 
$\sin ^2 \theta_W^{eff}$, $Z$ boson parameters and
couplings and translating them into the $M_W$,$M_t$
plane.  Right now, the three contours: $M_W$,$M_t$ from
the Tevatron; $M_W$ from LEP2; and the LEP, SLD, $\nu$N contour
are all consistent
with one another and tend favor a light Higgs mass.  
It is conceivable that the contours could all be 
consistent with the Standard Model yet inconsistent
with one another.  An inconsistency of this type
 would indicate non-Standard
Model physics.

The smaller the contours, the more stringent the
constraints on the Higgs boson mass and the 
Standard Model tests.  The goal of current and
future experiments is to measure electroweak 
parameters as precisely as possible to further
constrain and test the Standard Model.  Currently,
the crucial aspects of these measurements are
the top quark mass and the mass of the $W$ boson.

\subsubsection{The Measurement of $M_W$}

As stated previously, the dominant mechanism for 
$W$ boson production is quark-antiquark annihilation
($q\overline{q}^\prime \rightarrow W^\pm$).  The 
center of mass energy for this interaction, $\sqrt{\hat{s}}$
is much less than the $p\overline{p}$ center of mass energy
of $\sqrt{s}=1.8\, \rm TeV$.  This production mechanism 
leads to two important
consequences:
\begin{enumerate}
\item The energies of the annihilating 
 quark and antiquark are not  equal, meaning
the $W$ will be produced with a momentum component 
along the beam line ($p_z^W$).  Another way to put this
is to say that center-of-mass of the parton-parton
collision is moving in the lab frame.  The momentum of
the partons transverse to the beam direction is effectively
zero, so this center-of-mass motion is along the beam
direction.
\item Since the remnants of the $p$ and $\overline{p}$
carry a large amount of energy in the far forward 
direction (along the beam line) it is not possible to 
accurately measure the $\hat{s}$ of the interaction.
Therefore the initial $p_z$ of the  $W$ is not known. 
\end{enumerate}

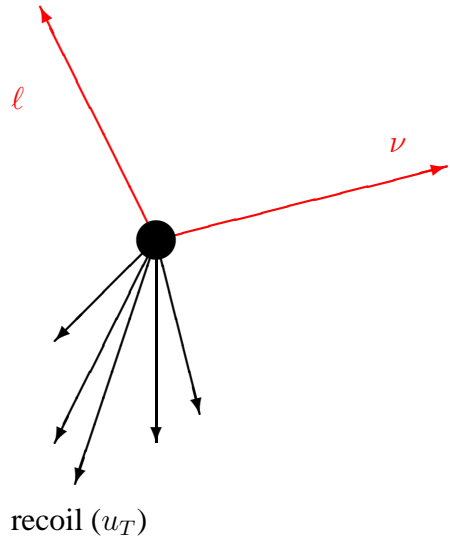
\begin{figure}[!th]
\begin{center}
\setlength{\unitlength}{1.1pt}
\begin{picture}(150,180)(0,-100)
\cred
\thicklines
\put (50,0){\vector(4,1){100}}
\put (130,30){\shortstack{$\nu$}}
\put (50,0){\vector(-1,2){40}}
\put (00,45){\shortstack{$\ell$}}
\cbla
\put (50,0){\vector(-1,-3){28}}
\put (50,0){\vector(-1,-2){35}}
\put (50,0){\vector(-1,-1){35}}
\put (50,0){\vector( 0,-1){70}}
\put (50,0){\vector(1,-4){15}}
\put (00,-100){\shortstack{recoil ($u_T$)}}
\put (50,0){\circle*{40}}
\end{picture}
\setlength{\unitlength}{1pt}
\caption{A cartoon of a $W\rightarrow \ell \nu_\ell$ decay.
The lepton is measured directly.  The transverse momentum
of the neutrino is inferred by the recoil energy ($u_T$).}
\label{fig:mwmeas}
\end{center}
\end{figure}

\begin{figure}[ht]
\begin{center}
\epsfxsize=3.5truein 
\hspace*{0.2truein}
\epsfbox{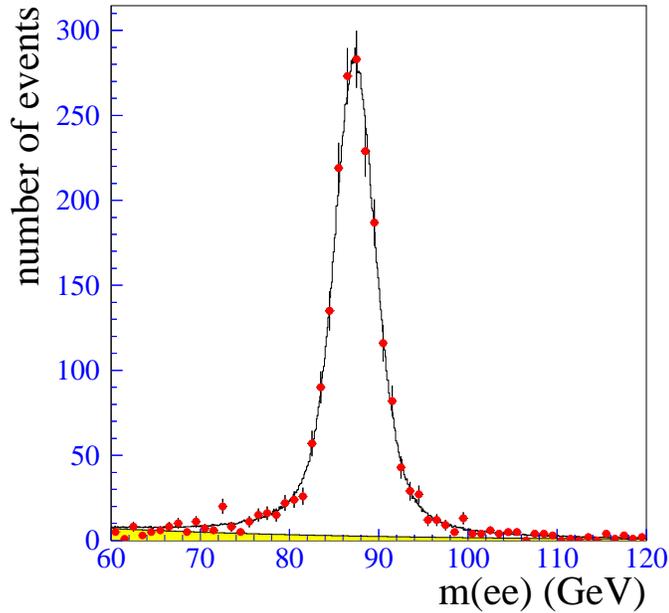}
\caption{The $Z^0$ mass as reconstructed in the mode
$Z^0\rightarrow e^+e^-$ by the D\O\ detector.  The
shaded region at the bottom of the plot is the background
contribution.  The peak does not fall exactly on the
true value of $M_Z$ because not all of the energy
corrections have been applied to the data.}
\label{fig:zmass}
\end{center}
\end{figure}

Because of these points, it is not possible to measure
the mass of the $W$ boson based upon the collision
energy, $\sqrt{\hat{s}}$.  We must measure the $W$ mass
by reconstructing the decay products.  

Recall that we are dealing with $W\rightarrow e\nu_e$ 
and $W\rightarrow \mu \nu_\mu$ modes.
The quantities associated with these decays that
 we can directly measure are:
\begin{itemize}
\item The momentum  of the muon,
$\vec{p}_\ell$.
\item The recoil energy, $\vec{u}$.
\end{itemize}

The lepton momentum can be measured in three dimensions.
The recoil energy can be measured in three dimensions, 
but since we do not know the initial $p_Z$ of the 
center of mass, the $z$ component of $\vec{u}$ and
$\vec{p}_\ell$ are 
of no use to us.   Since we know that (to very good
approximation) $p_x^W + u_x = p_y^W + u_y  = 0$, we can implement
conservation of momentum in the transverse $(x,y)$ plane
and infer the transverse momentum of the neutrino.
Since we do not know $p_z^W$, we can not infer the $p_z^\nu$
from momentum conservation.
Even with three-dimensional measurements of $\vec{u}$ and
$\vec{p}_\ell$, it is not possible to 
unambiguously determine the 
neutrino momentum in three dimensions.  If it were possible
to determine $\vec{p}_\nu$,
then we could simply calculate the invariant mass of
the $\ell$-$\nu_\ell$ and measure the $W$ mass from
the resonance.

The case of $Z$ production as discussed above is
quite similar to $W$ production.  The difference, however,
is that the $Z$ can  decay to two charged leptons that
we can measure in the detector.  Figure~\ref{fig:zmass}
shows the reconstructed $Z$ mass in the mode $Z^0\rightarrow
e^+e^-$ from the D\O\ detector.  The $Z$ peak is clear
and well-resolved, with small backgrounds.

In the case of the $W$ mass, the information we have
is momentum of the lepton 
$\vec{p}^{\, \ell}$ and transverse momentum of
the neutrino, $\vec{p}_T^{\, \nu}$, which was inferred
from the transverse momentum of the lepton and
the transverse recoil energy ($\vec{u}_T$).

From the transverse momenta of the lepton and the
neutrino, we can calculate a quantity known as
the ``transverse mass'': 
\begin{displaymath}
M_T^W = \sqrt{2 p_T^\ell p_T^\nu (1 - \cos\Delta\phi_{\ell,\nu})},
\end{displaymath}
where $p_T^\ell$ and $p_T^\nu$ are the magnitudes
of the lepton and neutrino transverse
momenta and $\Delta\phi_{\ell,\nu}$
is the opening angle between the lepton and neutrino
in the $x,y$ plane.

The transverse mass equation may look familiar.
If we have two particles where we have measured the
momenta in
$3$ dimensions with momenta $\vec{p}_1$ and $\vec{p}_2$, 
then the invariant
mass of those two particles in the approximation that
the particles are massless is:
\begin{displaymath}
M_{1,2} = \sqrt{2 p_1 p_2 (1 - \cos\alpha)},
\end{displaymath}
where $\alpha$ is the opening angle (in $3$-dimensions) between the
two particles.

\begin{figure}[t]
\begin{center}
\epsfxsize=4.5truein 
\hspace*{0.4truein}
\epsfbox{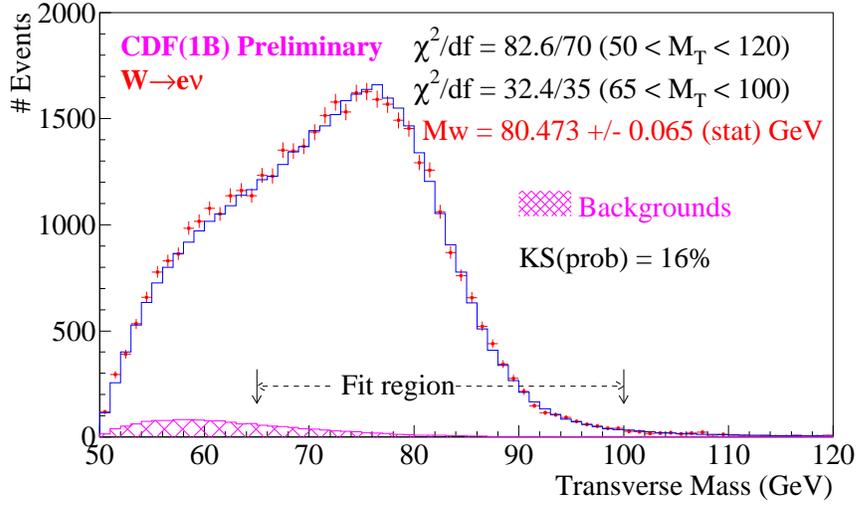}
\caption{The $W$ transverse mass in the mode $W\rightarrow e\nu_e$
as measured by CDF.  The points are the data, the histogram is the
fit.  The hatched region shows the background
contribution.}
\label{fig:wmass}
\end{center}
\end{figure}

By comparing the two equations, we can see that the
term ``transverse mass'' is accurate in that the calculation
is identical to the invariant mass except only the transverse
quantities are used.   If the $W$ boson has $p_Z^W = 0$,
then the transverse mass is exactly the invariant mass.  If
the $W$ boson has $|p_z^W|>0$, then the transverse mass is 
less than the invariant mass.  A $W$ boson transverse mass distribution
is shown in Fig.~\ref{fig:wmass}.

Although not quite as clean as a full invariant mass, the 
transverse mass distribution quite clearly contains information
about the $W$ mass.  By fitting this distribution, it is 
possible to extract a precise measurement of the $W$ mass.
There are three basic ingredients that determine the 
shape of the transverse mass distribution:
\begin{itemize}
\item $W$ boson production and decay.  
\item $p_T^\ell$ measurement.  
\item $u_T \Rightarrow p_T^\nu$ measurement. 
\end{itemize}
Each of these items will be discussed in detail below.
All of the details are ultimately combined
into a fast Monte Carlo simulation that is able to 
generate transverse mass spectra corresponding to various
values of the $W$ mass.  The measured transverse mass
distribution is then fit to the generated spectra and
the $W$ mass is extracted from this fit.

In the following subsections, we discuss each of the
elements required for precise $W$ mass determination.

\subsubsection{$W$  boson production and decay}

Modeling of the $W$ boson production and decay
 includes the
Breit-Wigner lineshape, parton
distribution functions, the momentum spectrum of the $W$ 
boson, the recoiling system and radiative corrections.
The intrinsic width of the $W$ boson is about $2.1\, \rm GeV/c^2$
which must be included in the fit.  The parton distribution
functions (PDF) are representations of the distributions of
valence quarks, sea quarks and gluons in the proton.  The
probability for specific processes as a function of $\hat{s}$
depend upon these distributions.  Related to the PDFs and the
production diagrams is the momentum distribution of the 
produced $W$ bosons.  The model of the recoil system must be
accurate.  Higher order QED 
diagrams, such as $W\rightarrow \ell \nu \gamma$
 are also included in the
modeling. 

\subsubsection{$p_T^\ell$ measurement}  

This aspect is quite crucial in the $W$ mass determination.
For muons, the transverse 
momentum is measured by the track 
curvature in the magnetic field.
For electrons, it is more accurate to measure the energy
(and infer the momentum) in the calorimeter because the resolution
is better and bremsstrahlung tends to bias the tracking measurement
of the curvature.  

The energy
scale is crucial. If we measure a muon with a 
transverse momentum of $30\, \rm GeV/c$, is the true momentum
$30\, \rm GeV/c$? Is it $29.9\, \rm GeV/c$?  Is it $30.1\, \rm GeV/c$?
Also, the resolution is important to understand.  For a 
measured momentum of $30\, \rm GeV/c$, we also need to know
the uncertainty on that value, because it will smear
out the transverse mass distribution.  In reality, the 
resolution is a rather small effect, much smaller than the
overall momentum scale.

To set the momentum/energy scale, we use ``calibration''
samples.  The $J/\psi$, $\Upsilon$ and $Z^0$ masses
are all known very precisely based upon measurements from
other experiments.  We can measure these masses using
$\mu^+\mu^-$ and $e^+e^-$ final states to calibrate 
our momentum scale.  If a muon measured 
with $p_T=29.9\, \rm GeV/c$
is truly  a muon with $p_T=30.0\, \rm GeV/c$, then we will
measure an incorrect $Z^0$ mass.  This scale can be noted and
ultimately corrected.

The $Z^0$ is particularly important for the $W$ mass measurement
because both its mass and the production mechanism are very
similar to that of the $W$.  They are not identical, though,
because the $Z^0$ is $10.7\, \rm GeV/c^2$ more massive
than the $W$.  Also, due to coupling and helicity 
considerations, the decay distributions are not identical
between the two.  They are quite close, however, and the
$Z^0$ provides a crucial calibration point.  The limiting
factor then arises from the number of $Z^0$ decays available.
As noted earlier, the ratio of observed leptonic $W$ decays
to $Z$ decays (${\cal R}$) is about 10:1.  In some cases, 
the limiting factor on the systematic uncertainty arises
from the statistics of the $Z$ samples.

\subsubsection{$u_T \Rightarrow p_T^\nu$ measurement} 

The recoil energy is required to infer the transverse
momentum of the neutrino.  Since the recoil energy is
largely hadronic and contains both charged and neutral 
components, it must be  measured with the calorimeter.
All of the 
charged and neutral energy recoiling against the $W$ 
is included in the measurement, so all sources of
calorimetric energy must be included in the model.  The
recoil distribution is affected by the collider environment, 
the resolution of the calorimeter, the coverage of the 
calorimeter  and the ability to separate
$u_T$ from $p_T^\ell$.  At typical Tevatron luminosities,
there are more than one, sometimes as many as six $p\overline{p}$
interactions per beam crossing.  Most of these are inelastic
events that have low transverse momentum.  However, there is no
way to directly separate out the contributions from other
interactions from the contributions of the $W$ recoil.  Instead,
this must be modeled and the background level subtracted
on an average basis.   Uncertainty in this background subtraction
leads to uncertainty in $M_W$.

The
hadronic energy resolution of the calorimeter is much larger 
(\it i.e. \rm worse)
than the resolution on the lepton energy.  Therefore,
the resolution on the neutrino $p_T$ is determined by the
hadronic energy resolution.  The smaller this resolution,
the less smeared the transverse mass distribution.

The coverage of the calorimeter must be understood, also,
because some of the recoil can be carried away at very
small angles to the beamline, where there is no instrumentation.

Finally, the recoil measurement is a sum of all calorimeter 
energy except the energy of the lepton.  In the case of the
muon channel, it is pretty straightforward to subtract the
contribution from the muon.  For the electron, some
of the recoil energy is included in the electron energy
cluster in the calorimeter simply because the recoil and
electron energy ``overlap''.  This affects both the electron
energy measurement and the $u_T$ measurement and therefore
we must correct for that effect.

\begin{figure}[thb]
\begin{center}
\epsfxsize=4.0truein 
\hspace*{0.4truein}
\epsfbox{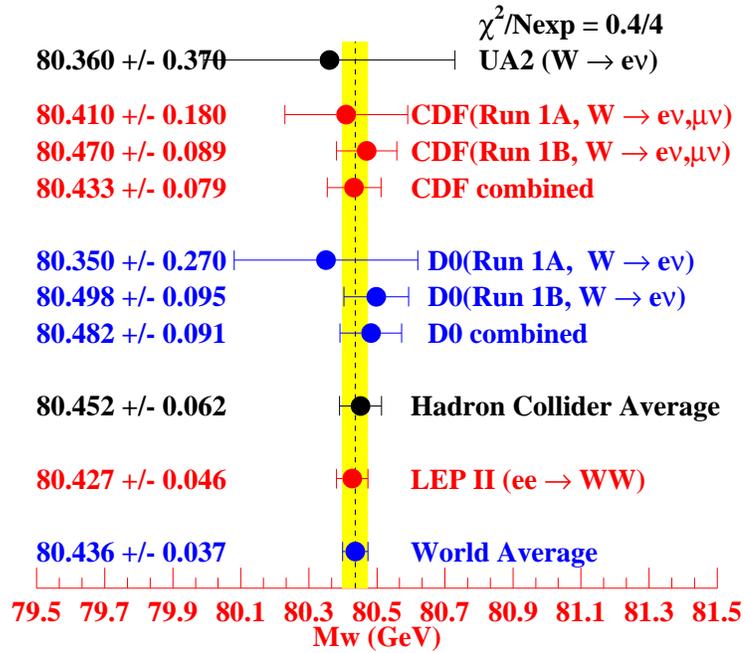}
\caption{Summary of direct measurements of the $W$ mass.
The LEP II point is the combination of four experiments, 
while the CDF and D\O\ results are shown separately.
The world average uncertainty is $37\, \rm MeV$.}
\label{fig:wworld}
\end{center}
\end{figure}

\subsubsection{$W$ Mass Summary}

Each of these pieces needs to be fully and accurately modeled
in order to understand how they effect the transverse mass
distribution.   There are many important aspects to this
analysis, but the most important is the lepton
energy scale.  A great deal of work has gone into calibrating,
checking and understanding the lepton energy scale.

Details of the D\O\ and CDF $W$ mass measurements may
be found in Refs~\cite{d0wmass,cdfwmass}.
For a recent compilation of the world's $W$
mass measurements may be found in Ref~\cite{allwmass}.

\begin{figure}[h!]
\begin{center}
\epsfxsize=4.20truein 
\hspace*{0.4truein}
\epsfbox{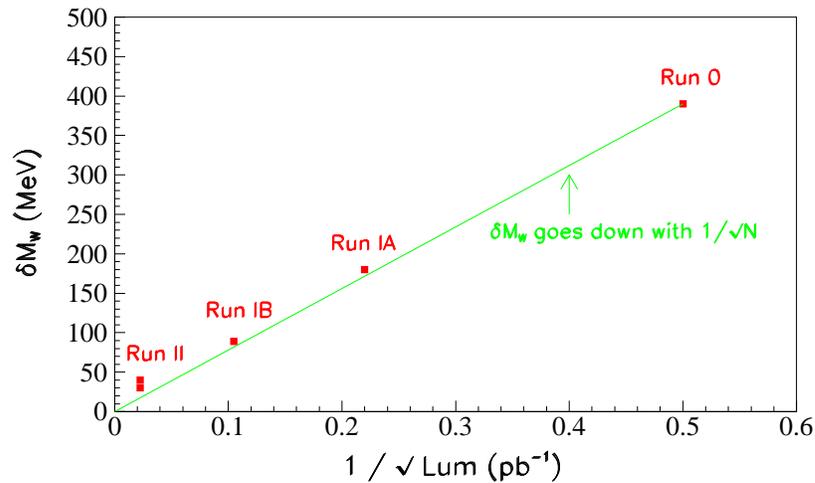}
\label{fig:werr}
\caption{The $W$ mass uncertainty as a function of 
data size.  Both the statistical and systematic errors have
continually fallen linearly as $1/\sqrt{N}$.  This trend 
will continue in the future, although the ultimate Run~II
sensitivity will deviate from the line.
}
\end{center}
\end{figure}

\begin{figure}[h!]
\begin{center}
\epsfxsize=4.20truein 
\hspace*{0.4truein}
\epsfbox{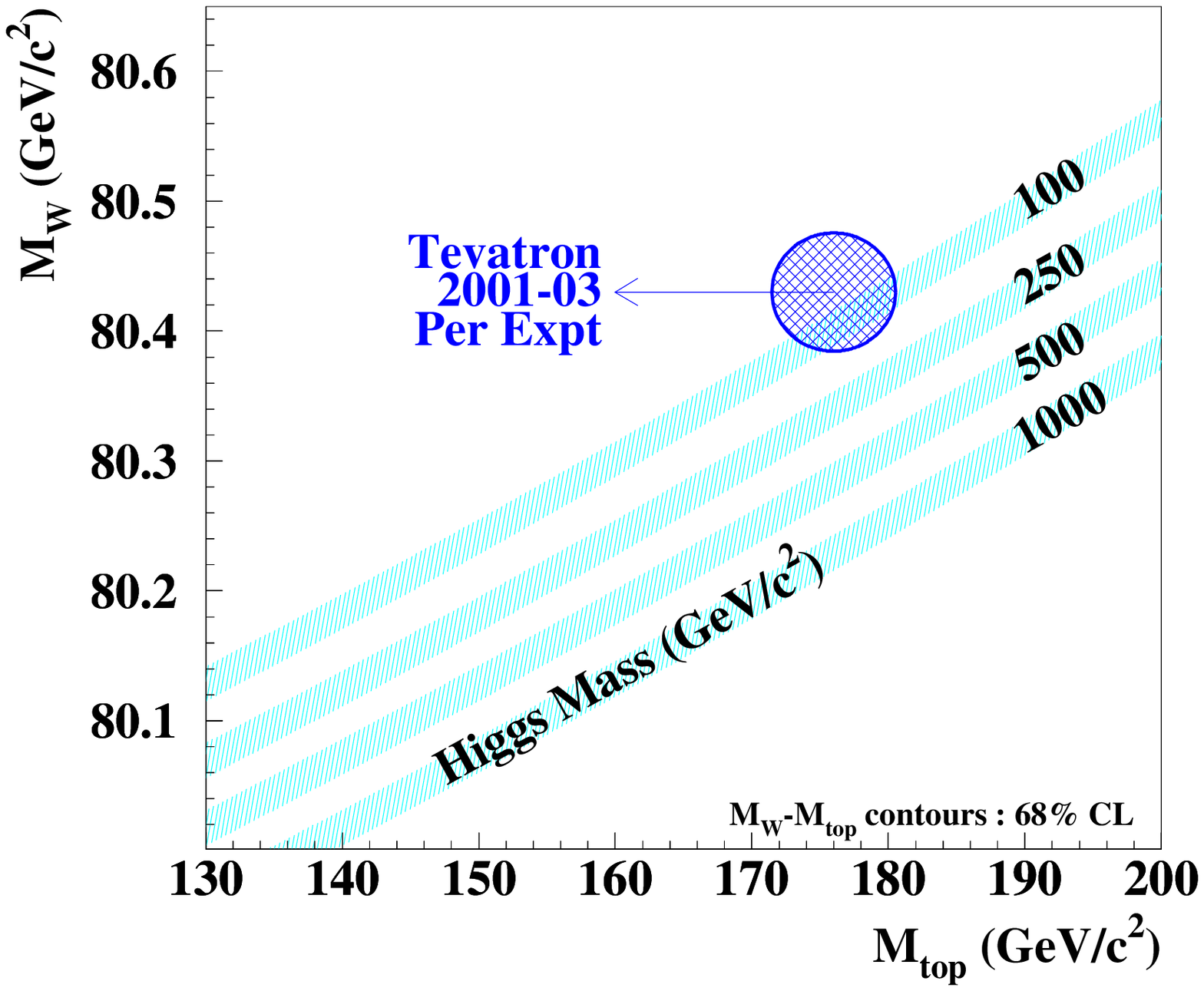}
\caption{The $M_W$ versus $M_t$ plot at the end of
Run~II.  The central value is plotted as it currently 
stands.  For Run~II, we anticipate $\delta M_W \sim\! 30\, \rm MeV$
and $\delta M_t \sim \! 3\, \rm GeV$.  These results will further
test and constrain the Standard Model.}
\label{fig:mwmt2}
\end{center}
\end{figure}

\subsubsection{The Future} 

  In addition to the Tevatron upgrades
for Run~II, the  D\O\ and CDF collaborations 
are significantly
upgrading their detectors \cite{d02,cdf2}.
Figure~\ref{fig:mwmt2} shows how the uncertainty
on the $W$ mass has progressed over time.   
Since $N\propto {\cal L}_{int}$, the horizontal axis, plotted
as $\sqrt{{\cal L}_{int}}$ is equivalent to $\sqrt{N_W}$, with
$N_W$ being the number of identified $W$ boson decays.  So
far, the uncertainty on the $W$ mass has fallen linearly
with $1/\sqrt{N_W}$.  We expect the statistical uncertainty
to fall as $1/\sqrt{N_W}$.  The recent measurements of $M_W$
are not dominated by the statistical uncertainty, however.
To maintain the $1/\sqrt{N_W}$ behavior of the total 
error, both the systematic and statistical uncertainties
must fall as the statistics increase.
This can be understood from the fact that
many of the systematic uncertainties are limited by the
statistics of the control samples, such as $Z^0\rightarrow \ell^+
\ell^-$.  As those samples grow, the systematic uncertainties
fall.  

Tevatron Run~II is projected to move slightly away from
the strict $1/\sqrt{N_W}$ behavior as some of the systematic
uncertainties become limited by factors other than the
statistics of the control samples.  Nevertheless, the uncertainty
is expected to be significantly reduced.  The combined 
$W$ mass uncertainty from D\O\ and CDF is expected to be
between $20$ and $40\, \rm Mev/c^2$ in Run~II.

At the same time, the uncertainty on the top quark mass
will also be reduced.  Figure~\ref{fig:mwmt2} also  shows
what the $M_W,M_t$ plot could look like by about 2003.
For this plot, we assume that the central measured value
is the same as it is currently, simply to demonstrate 
how the uncertainty contours will look at that time.
This compares quite favorably to the current version of
this plot, shown previously in Fig.~\ref{fig:mwmt}.

\section{$B$ Physics Results}

Since the first observation of a violation of 
charge-conjugation parity ($CP$) invariance  
in the neutral kaon system in 1964,\cite{cronin} 
there has been an ongoing effort to further 
understand the nature of the phenomenon.
To date, violation of $CP$ symmetry  has not
been directly observed anywhere other than
the neutral kaon system.  
Within the framework of the Standard Model, 
$CP$ violation arises from a complex phase
in the 
Cabibbo-Kobayashi-Maskawa (CKM) quark 
mixing matrix,\cite{ckm} although  the physics
responsible for the origin of this phase is 
not understood.
The goal of current and future measurements in the
$K$ and $B$ meson systems is to continue to improve
the constraints upon the mixing
matrix and further test the Standard Model.
Inconsistencies would  point towards physics
beyond the Standard Model.

In recent years, the importance and experimental advantages
of the $B$ system have been emphasized.\cite{sanda}  The long lifetime of
the $b$ quark, the large top quark mass  and the 
observation of $B^0/\overline{B^0}$ mixing with a long
oscillation time all conspire to make the $B$ system
fruitful in the study of the CKM matrix.  Three 
$e^+e^-$ $B$-factories running on the $\Upsilon$(4s) 
resonance in
addition to experiments at HERA and the Tevatron indicate the
current level of  interest and knowledge to be gained by
detailed study of the $B$ hadron decays.

This section is an introduction to $CP$ violation
in the $B$ system, with a focus on experimental issues. 
After a some notational definitions, I will give a
brief overview of the CKM matrix and $B^0/\overline{B^0}$
mixing.  Following that,  
I  will discuss experimental elements of flavor tagging,
which is a crucial component in mixing and 
$CP$ asymmetry measurements.
Our discussion of $CP$ violation in the $B$ system
will be presented in the framework of the specific example of 
the measurement of $\sin 2 \beta$ using $B^0/\overline{B^0}\rightarrow
J/\psi K^0_S$ decays by the CDF Collaboration.   Finally, I
will briefly survey  future measurements.

\subsection{Notation}
There are enough $B$'s and $b$'s associated with this topic that it
is worthwhile  to specifically spell out our notation.
First of all, we will refer to bottom (antibottom) quarks
using  small letters: $b$ ($\overline{b}$).  When we
are referring to generic hadrons containing a bottom quark
(\it e.g. \rm $|b\overline{q}>$, where $q$ is any quark type),
we will use a capital $B$ with no specific subscripts or superscripts.

In the cases where we are referring to specific bottom mesons
or baryons, we will us the notation listed in Table~\ref{ta:b}.
Neutral $B$ mesons follow the convention
of the neutral kaon system, where $K^0 = |\overline{s}d>$
and $\overline{K^0} = |s\overline{d}>$.

\begin{table}[hbt]
\caption{$B$ mesons and baryons.  This is an incomplete 
list, as there are excited states of the mesons and baryons
(\it e.g. \rm ${B^*}^0$).  Also, a large number of $B$-baryon
states 
are 
not listed (\it e.g. \rm $\Sigma^-_b = | ddb >$).}
\label{ta:b}
\begin{center}
\begin{tabular}{lll}
name & $\overline{b}$ hadron & $b$ hadron  \\
\hline 
charged $B$ meson &
$B^+ =|\overline{b}u>$ \hspace*{0.5truein} &
       $B^- = |b\overline{u}>$ \\
neutral $B$ meson & 
$B^0 = |\overline{b}d>$ & $\overline{B^0} = |b\overline{d}> $\\
$B_s$ ($B$-sub-$s$) meson\ \ \ \ \ \  \   & 
 $B^0_s = |\overline{b}s>$ &  $\overline{B^0_s}=|b\overline{s}>$\\
$B_c$ ($B$-sub-$c$) meson &  $B^+_c = |\overline{b}c>$ &  
$B^-_c = |b\overline{c}>$ \\  
$\Lambda_b$ (Lambda-$b$) &
  $\overline{\Lambda_b} = |\overline{u}
\overline{d}\overline{b}> $
& $\Lambda_b = |udb> $ \\
$\Upsilon$ (Upsilon) & \multicolumn{2}{c}{$\Upsilon = |\overline{b}b>$}  \\
\hline
\end{tabular}
\end{center}
\end{table}


\subsection{Overview: the  Cabibbo-Kobayashi-Maskawa  Matrix}

Within the framework
of the Standard Model, $CP$ nonconservation  arises through a non-trivial
phase in the Cabibbo-Kobayashi-Maskawa (CKM) quark mixing matrix.\cite{ckm}
The CKM matrix $V$ is the unitary matrix that
transforms  the mass  eigenstates into the weak
eigenstates:
\begin{eqnarray}
  V =&&\pmatrix{V_{ud}&V_{us}&V_{ub}\cr
                V_{cd}&V_{cs}&V_{cb}\cr
                V_{td}&V_{ts}&V_{tb}\cr}\\
 \simeq && \pmatrix{1-{\lambda^2\over2}&\lambda&A\lambda^3(\rho\!-\!i\eta)\cr
                      -\lambda&1-{\lambda^2\over2}&A\lambda^2\cr
                      A\lambda^3(1\!-\!\rho\!-\!i\eta)&-A\lambda^2&1\cr}
+ {\cal O}(\lambda^4).
\end{eqnarray}
The second matrix is a useful phenomenological 
parameterization of the quark mixing matrix suggested
by Wolfenstein,\cite{lincoln} in which $\lambda$ is the 
sine of the Cabibbo angle, $\lambda = \sin \theta_C \simeq 0.22$.
The CKM matrix is an arbitrary three-dimensional rotation
matrix.  The only requirement \it a priori \rm is that
it be  unitary -- the 
value of the elements can take on any value so long as unitarity
is preserved.  The Wolfenstein parameterization arose based upon 
experimental results indicating that the matrix is nearly diagonal.
Using experimental results on $V_{us}$ and $V_{cb}$ along with
the unitarity requirement, Wolfenstein proposed the commonly-seen
expansion shown here.

\begin{figure}[t]
\begin{center}
\epsfxsize=3.0truein 
\hspace*{0.4truein}
\epsfbox{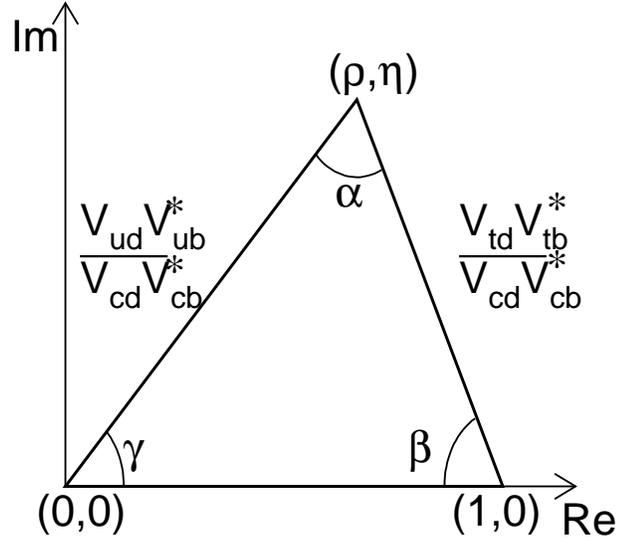} 
\caption{The unitarity triangle.  The horizontal axis
is the real axis; the vertical axis is the imaginary
axis.  The apex of the triangle is $(\rho,\eta)$.}
\label{fig:triangle}
\end{center}
\end{figure}

The condition of unitarity, $V^\dagger V = 1$, 
yields several relations, the most important of which is a relation between 
the first and third columns of the matrix, given by:

\begin{equation}
V^*_{ub}V_{ud} 
+ V^*_{cb}V_{cd}
+ V^*_{tb}V_{td}= 0. 
\end{equation}
This relation, after division by  $V^*_{cb}V_{cd}$, 
is  displayed graphically in Fig.~\ref{fig:triangle} as a
triangle in the complex ($\rho$-$\eta$) plane, and is 
known as the unitarity triangle.\cite{linglee}
$CP$ violation in the Standard Model 
manifests itself as  a nonzero value of $\eta$, the height of the triangle,
which indicates the presence of an imaginary CKM component.

The ``unitarity triangle'' is simply a graph of a single point in
the complex plane: $(\rho,\eta)$.  We use the triangle to show how
these two numbers are related to the CKM elements.  Different 
experimental measurements are sensitive to different aspects of
the unitarity triangle, \it i.e. \rm they are sensitive to different
combinations of $\rho$ and $\eta$.

Six unique triangles can be constructed
from unitary relations (six more are complex conjugates
of the first six.)  The one shown here is the most useful
because all of the sides are of ${\cal O}(\lambda)$, insuring that none
of the three interior angles is near $0^\circ$ or $180^\circ$.
The other triangles are ``squashed'' having one side ${\cal O}(\lambda^2)$ or
${\cal O}(\lambda^3)$ smaller than the other two sides.

The goal of current and future experiments
in the $K$ and $B$ system
is to measure as many aspects of the triangle as possible in as many
ways as possible.  Inconsistencies in these measurements will point
to physics beyond the Standard Model and hopefully give us some indication
from where these ``fundamental constants'' arise.

Based upon current measurements in the $K$ and $B$ system, such as
$B^0/\overline{B^0}$ mixing, $K\rightarrow \pi^\pm \ell^\mp \nu$,
$b \rightarrow u$ decays and $b\rightarrow c$ decays, the CKM solution
indicates that the $CP$ violating phase is large.  The fact that
$CP$ violation in the $K$ system is small, ${\cal O}(0.1\%)$, arises from
the fact that the magnitude of the matrix element $V_{td}$ is 
rather small.  An alternate solution would be if the $CP$ violating
phase were to be small and the magnitude of $V_{td}$ larger.
Direct measurements of $CP$ violation in the $B$ system will permit
clear distinction between the two cases.\cite{nir}

\subsection{$B^0/\overline{B^0}$ Mixing}

Mixing occurs in the neutral $K$ and $B$ systems because the
electroweak eigenstates and the strong interaction eigenstates
are not the same.  If we start with a $B^0$ meson,
then the probability that we will see a $B^0$($\overline{B^0}$)
at a given time, $t$, is
\begin{displaymath}
P(B^0(t)) =
       \frac{1}{2\tau}e^{-{t\over{\tau}}} (1  +
       \cos(\Delta m_d t)) 
\end{displaymath}
\begin{equation}
 P(\overline{B^0}(t)) =
       \frac{1}{2\tau}e^{-{t\over{\tau}}} (1  -
       \cos(\Delta m_d t))
\end{equation}
where $\tau$ is the $B^0$ lifetime and 
$\Delta m_d = m_H - m_L$,\footnote{The subscript $d$ on $\Delta m_d$
refers to the down quark in the neutral $B$ meson.  This is to distinguish
from the $B^0_s/\overline{B^0_s}$ mass difference, which is 
written as $\Delta m_s$ with the subscript $s$ referring to the
strange quark.} 
 where $m_H$ and $m_L$ are the masses of the 
heavy and light weak eigenstates of the mesons.   
The mass difference $\Delta m_d$ 
in the $B^0/\overline{B^0}$ system is  
relatively small, therefore the mixing frequency
is rather low.   In units where $\hbar=c=1$, the mass difference
is presented in units of $\rm ps^{-1}$.  The current world average
for $\Delta m_d$ is $0.487\pm0.014 \, \rm ps^{-1}$.\cite{lepwg}
With this mass difference, the oscillation period for 
$B^0/\overline{B^0}$ is close to nine $B$ lifetimes.

\begin{figure}[htb]
\begin{center}
\epsfxsize=25pc 
\hspace*{0.5truein}
\epsfbox{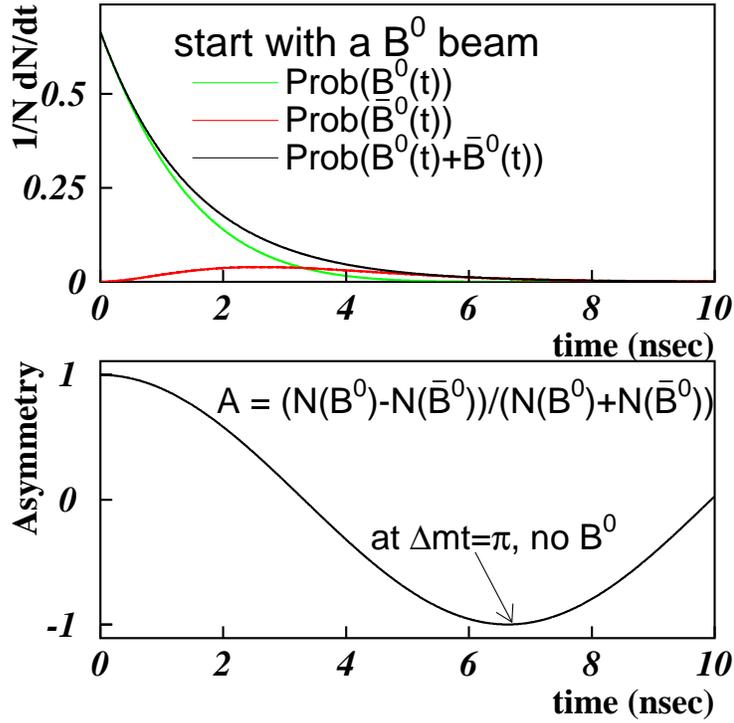} 
\caption{$B^0/\overline{B^0}$ mixing.  The top plot shows the 
probability functions for both $B^0$ and $\overline{B^0}$ as a 
function of time.  At $t=0$, we have 100\% $B^0$.  As time increases,
the mesons decay away exponentially $\propto e^{-{t\over{\tau}}}$,
but some of the $B^0$ mesons become $\overline{B^0}$ mesons.
The bottom plot shows the asymmetry so that the exponential effect
has been removed.  At $\Delta m_d t = \pi$ 
($\Rightarrow t \simeq 4.4\tau \simeq 6.8\, \rm ps 
 $, because the $B^0$ lifetime is $\tau = 1.56\, \rm ps$.), 
\bf all
of the remaining mesons are $\overline{B^0}$! \rm } 
\label{fig:mix}
\end{center}
\end{figure}

Mixing is shown graphically in Fig.~\ref{fig:mix}.
When we begin with a beam of $B^0$ mesons, they disappear
at a rate faster then $e^{-t/\tau}$, because some $B^0$ mesons
are decaying and some are oscillating into $\overline{B^0}$
mesons.  The sum of $B^0$ plus $\overline{B^0}$ decay at a 
rate $e^{-t/\tau}$.

\begin{figure}[t]
\begin{center}
\epsfxsize=24pc 
\hspace*{0.5truein}
\epsfbox{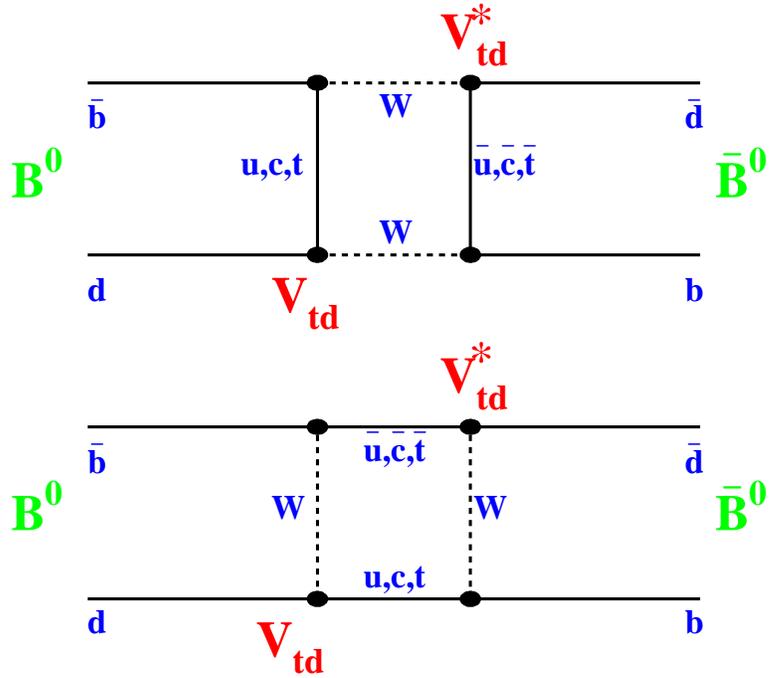}
\caption{$B^0/\overline{B^0}$ mixing diagrams.  The diagrams shown
are for $B^0$ oscillating into $\overline{B^0}$.  The charge-conjugate
process ($\overline{B^0}$ oscillating into $B^0$) 
takes place as well.   The top quark
dominates these $2^{nd}$ order weak transitions, which is why
$V_{td}$ (and not $V_{ud}$ or $V_{cd}$) is shown at the vertices.}
\label{fig:box}
\end{center}
\end{figure}

Mixing in the neutral $B$ system is a second order 
$\Delta B = 2$
transition\footnote{The $B$ used here refers to the 
``bottomness'' quantum number.  Since the box diagram is 
responsible for annihilating a $\overline{b}$ and producing
a $b$ (or vice versa) the change in the bottomness quantum
number is $\Delta B = 2$.}
that proceeds through ``box'' diagrams shown in 
Fig.~\ref{fig:box}.  All up-type quarks ($u$,$c$ and $t$)
are eligible to run
around in the box, but the heavy top quark dominates because
the amplitude is proportional to the mass of the fermion.  
As a 
consequence of this, there are two $top$-$W$-$down$ vertices ($V_{td}$)
in the
dominant box diagram.   This will play a roll in $CP$ violation that
we will discuss below.

The Feynman diagrams for $B^0_s/\overline{B^0_s}$ look quite
similar with the exception that
 the $top$-$W$-$down$ vertices ($V_{td}$) are replaced by the
$top$-$W$-$strange$ 
($V_{ts}$) vertices.  Since $|V_{ts}|>|V_{td}|$, the 
$B_s$ system oscillates much faster than does the $B_d$ system.
Put another way, 
$\Delta m_s$ is much
larger than $\Delta m_d$.  The $B_s$ oscillates so quickly
that the oscillation period is smaller than the experimental 
resolution on the decay time of the $B_s$.  In other words, we
can identify and  
distinguish between $B^0_s$ and $\overline{B^0_s}$
mesons at the time of decay, but the resolution of the decay
time is not yet good enough to resolve the oscillations.
The current experimental
bound is $\Delta m_s > 14\, \rm ps^{-1}$, which means that
the $B_s$ fully mixes in less than 0.17 lifetimes!

\subsection{Flavor Tagging}

To measure time-dependent mixing, it is necessary to know
what the flavor of the meson was at the time of production
and at the time of decay.  For example, an ``unmixed'' event
would be an event where a neutral $B$ meson was produced as
a $B^0$ and decayed as a $B^0$.  A ``mixed'' event would
be one where a neutral $B$ meson was produced as a $B^0$ 
but decayed as a $\overline{B^0}$.  Typically, mixing
results are plotted (bottom plot of Fig.~\ref{fig:mix})
as an asymmetry: 
${\cal A} = (N_{unmixed}-N_{mixed})/(N_{unmixed} + N_{mixed})$.  
This has the advantage of removing 
the exponential term from the decay probabilities.  Once
plotted in this way, the functional form of the mixing
is ${\cal A} = \cos\Delta m_d t$.\footnote{Another common
way to display mixing data is of the form ${\cal A} =
N_{mixed}/(N_{unmixed}+N_{mixed})$ which then takes
the functional form ${\cal A} = 1- \cos \Delta m_d t$.} \

Experimentally, the determination of the flavor of the
$B$ meson at the time of production and/or the time of
decay is referred to as ``flavor tagging".
Flavor tagging is an inexact science.  The $B$ mesons have numerous
decay modes, thanks in large part to the large phase space
for production of light hadrons in the dominant $B\rightarrow D
\rightarrow X_s$ decay, where $D$ and $X_s$ represent generic
charmed and strange hadrons respectively.  There is very low
efficiency for fully reconstructing $B$ states.  Therefore more 
inclusive techniques must be used to attempt to identify flavor.

Since flavor tagging is imprecise, it is crucial that 
we measure  our success/failure rate.  There are two parameters
required to describe flavor tagging.  The first is known as the
tagging efficiency, $\epsilon$, which is simply the fraction of
events that are tagged.  For example, if we are only able to
identify a lepton on $10\% $ of all of the events in our sample, 
then the lepton tagging
efficiency is $10\%$.  We can not distinguish a $B^0$ from a 
$\overline{B^0}$ in the other $90\%$ of the events because there
was no lepton found to identify the flavor.

The second parameter is associated with how often the
identified flavor is correct.  A ``mistag'' is an event
where the flavor was classified incorrectly.  A mistag rate ($w$)
of 40\% is not unusual; while a mistag rate of 50\% would mean
that no flavor information is available -- equivalent
to flipping a coin.  Another way to classify the success rate
is through a variable called the ``dilution'' (${\cal D}$), defined as
\begin{equation}
{\cal D} = {N_{right} - N_{wrong}\over{N_{right} + N_{wrong}}} = 1-2w
\end{equation}
where $N_{right}$($N_{wrong}$) are the number of events tagged
correctly (incorrectly).   
The term is dubbed ``dilution'' because
it dilutes the true asymmetry:
\begin{equation}
{\cal A}_{observed} = {\cal D}{\cal A}_{true}
\end{equation}
where ${\cal A}_{observed}$ is the experimentally measured 
asymmetry and ${\cal A}_{true}$ is the measurement of 
the real asymmetry we are
trying to uncover.\footnote{The choice of the term ``dilution''
here is unfortunate, since in this case a high dilution is
good and a low dilution is bad.  The definition comes about
because the factor ${\cal D} = 1-2w$ ``dilutes'' the measured
asymmetry.  If our flavor tagging algorithm were perfect (no 
mistags) then we would have ${\cal D} = 1$, the highest possible
dilution.}

In the following subsections 
we discuss some commonly used flavor tagging techniques.
The methods outlined below are all utilized in mixing
analyses.  However, it is the initial state flavor tag
that is important for $CP$ asymmetry measurements.
The methods discussed here are summarized in Table~\ref{ta:tag}.

\begin{table}[t]
\caption{Methods of flavor tagging.  These methods can be used in
mixing analyses as well as $CP$ asymmetry measurements.  In the
case of $CP$ asymmetry measurements, the initial state flavor is
the one of interest, as will be shown later.}
\label{ta:tag}
\begin{center}
\begin{tabular}{|l|l|}
\hline
method & initial/final state tag \\
\hline
exclusive reconstruction &  final  \\
partial reconstruction &  final\\
lepton tagging & initial/final\\
jet charge tagging &  initial\\
same side tagging & initial\\
\hline
\end{tabular}
\end{center}
\end{table}

\subsubsection{Full/Partial Reconstruction}
The flavor of the $B$ meson at the time of decay
can be determined from the decay products.  An
example of this is $B^0\rightarrow D^- \pi^+$, with
$D^- \rightarrow K^+\pi^-\pi^-$.  This all-charged
final state is  an unambiguous signature of a
$B^0$ meson at the time of decay.
The drawback of the full reconstruction technique is
that both the branching ratios to specific final states 
and  reconstruction efficiencies are low.

To improve upon this, we can relax by performing
a ``partial'' reconstruction.  An example of this
relating to the example above is to reconstruct
$B^0\rightarrow D^- X$, with $D^- \rightarrow K^+\pi^-\pi^-$.
In this case, the $X$ would include the state listed 
above, but
would include all other decays of this type (\it e.g. \rm
$B^0 \rightarrow D^- \pi^+ \pi^0$.)  Partial reconstruction
is not as clean as full reconstruction.  Since
it is also possible to have $\overline{B^0}\rightarrow
D^- X$, $B^+ \rightarrow D^- X$,
$B^0_s \rightarrow D^- X$ in addition to direct charm production,
where $\overline{c}\rightarrow D^-$.  Therefore 
the reconstruction of a 
$D^-$ meson is not an unambiguous signature for a $B^0$ meson.  
These other contributions  must
be accounted for in the extraction of $\Delta m_d$.

\subsubsection{Initial State Tagging}

\begin{figure}[t]
\begin{center}
\epsfxsize=32pc 
\hspace*{0.25truein}
\epsfbox{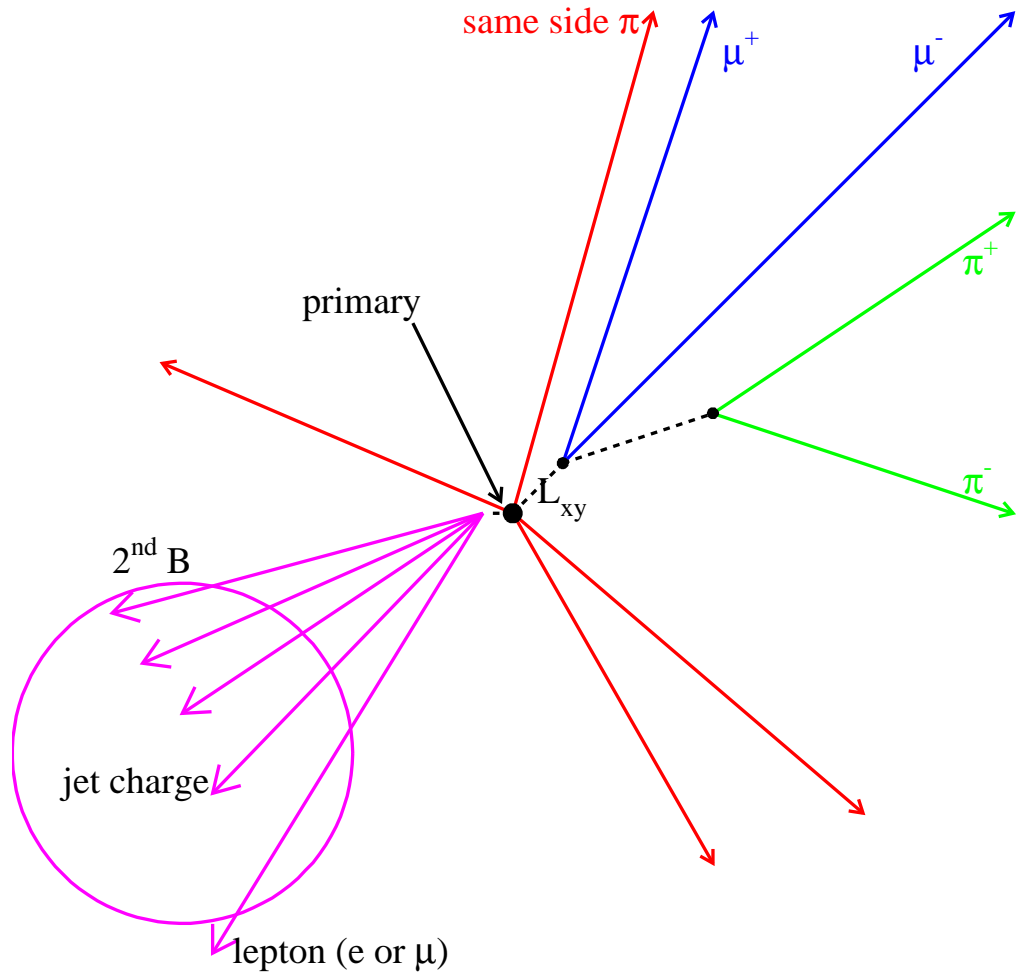}
\caption{Initial state flavor tags.  This example shows a 
reconstructed $J/\psi K^0_S$ final state.  The other information
in the event is used to identify the flavor of the $B^0$ or
$\overline{B^0}$ at the time of production.}
\label{fig:tags}
\end{center}
\end{figure}

It is not
possible to 
measure the flavor of a neutral 
$B$ meson at the time of production using full or 
partial reconstruction, because the decay only reflects
the flavor of the final state.  To perform initial
state flavor tagging, two types of methods are employed.
The first technique, known as opposite-side tagging,  
involves looking at the other $B$ hadron in the
event.  The second technique, known as same-side tagging, 
involves looking at the local correlation
of charged tracks near the $B$.

\begin{figure}[t]
\begin{center}
\epsfxsize=34pc 
\epsfbox{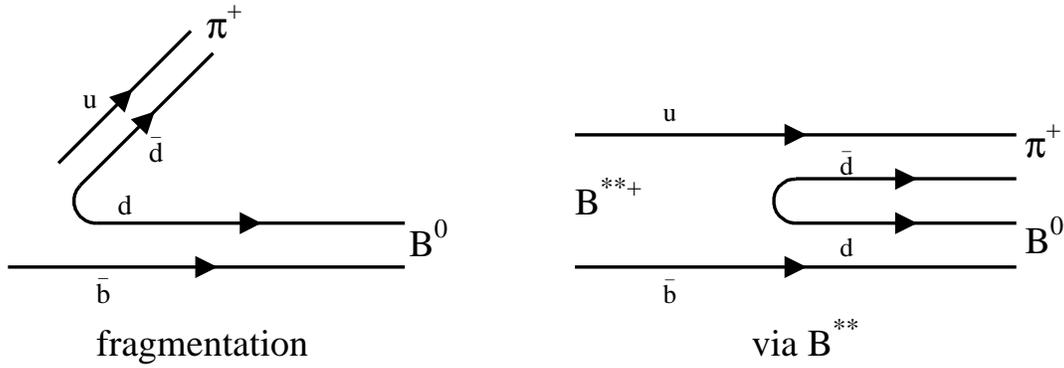}
\caption{Same-side flavor tagging.  In both cases shown above,
a $\overline{b}$ quark is produced and ultimately ends up
as a $B^0$ meson.  In the left diagram, the $\overline{b}$ quark
has grabbed a $d$ quark from the vacuum.  The remaining $d$
quark has paired with a $\overline{u}$ quark to make a $\pi^+$.
In the right diagram, the $\overline{b}$ quark grabs a $u$ quark
to produce a radially-excited ${B^{**}}^+$ state.  The ${B^{**}}^+$
then decays to a $B^0\pi^+$.  In both cases, the $\pi^+$ is 
associated with a $B^0$ meson and a $\pi^-$ would be associated
with a $\overline{B^0}$ meson.  No information about the other
$B$ hadron in the event is required. }
\label{fig:sst}
\end{center}
\end{figure}

In the case of opposite side tagging, we are taking advantage
of the fact that $b$ and $\overline{b}$ are produced in pairs.
If we determine the flavor of one $B$ hadron, we can infer the
flavor of the other $B$ hadron.  This is not perfect, of course,
because in addition to the complications mentioned above,
the opposite side $B$ hadron may have been a $B^0$ or $B^0_s$ 
and mixed.

Three types of opposite tagging are commonly used:
\begin{itemize}
 \item \bf lepton tagging: \rm  identify $B\rightarrow \ell \nu X$.  
The lepton carries the charge of the $b$.
 \item \bf kaon tagging: \rm identify $B\rightarrow D \rightarrow K$ 
($b\rightarrow c\rightarrow s$).  The strange particle carries the
charge of the $b$.
\item \bf jet charge tagging: \rm  identify a ``jet'' associated with
$B\rightarrow X$ and perform a momentum weighted charge sum.  On
average, the net charge of the jet will reflect the charge of the
$b$.
\end{itemize}
Each of the methods has different experimental requirements
and therefore different sets of  positive and negative aspects.  
For example,
with lepton tagging, the branching ratio and efficiency are rather
low.  In addition, there are mistags that come from $b\rightarrow c
\rightarrow \ell$.  On the other hand, lepton tags tend to have
high dilution ($=$large ${\cal D}$).  
For jet charge tagging, the dilution is lower $=$small ${\cal D}$), 
but
we are more likely to find a jet, which means a higher tagging
efficiency.

By contrast, same-side tagging makes no requirement on the second
$B$ hadron in the event.  It instead takes advantage of the 
effects associated with hadronization.  When a $\overline{b}$ quark
becomes a $B^0$ meson, it must pair up with a $d$ quark.  Since
quark pairs pop-up from the vacuum, there is a $\overline{d}$ quark
associated with the $d$ quark.  Now if the $\overline{d}$ quark
grabs a $u$ quark, then there is a $\pi^+$ associated with the
$B^0$.   This is shown in Fig.~\ref{fig:sst}.  An alternative
path to the same correlation is through the production of a 
$B^{**}$ state.  In either case, the correlation is: $B^0\pi^+$
and $\overline{B^0}\pi^-$.  In our example above, if the
$\overline{d}$ grabs a $d$ quark, then we have a $\pi^0$, in 
which case the first-order correlation is lost.

The same-side technique has the advantage of not relying on the
second $B$ hadron in the event.  The disadvantage is that, 
depending upon the hadronization process for a given event, 
the measured correlation may be absent
or may be of the wrong sign.   For example, the 
correlation would not be measurable if the mesons from
the fragmentation chain were neutral.  If the up quark in
Fig.~\ref{fig:sst} were replaced by a down quark, then the
associated meson would be a $\pi^0$.  Likewise, wrong-sign
correlations are present: if the up quark in Fig.~\ref{fig:sst}
were replaced with a strange quark, then a ${K^*}^0$ would
be produced, with ${K^*}^0\rightarrow K^-\pi^+$.  If the $K^-$
is selected as the tagging track, then the wrong-sign is 
measured.  This type of mistag can be reduced through the 
use of particle identification to separate charged kaons, 
pions and protons.

As will be seen below, initial state flavor tagging is 
a crucial aspect in measuring $CP$ asymmetries in the $B$
system.  In the analysis we will discuss here, three of 
the four initial state tagging methods are used: lepton tagging,
jet-charge tagging and same-side tagging.

\subsection{$CP$ Violation Via Mixing}

For Standard Model $CP$ violation to occur, 
we need an interference to expose the complex CKM
phase.
The $CP$ violating phase in $V_{td}$ can manifest itself
through the $\Delta B=2$ box diagrams responsible for 
$B^0/\overline{B^0}$ mixing. 
In the Standard Model, the decay
mode  $B^0/\overline{B^0}
\rightarrow J/\psi K^0_S$ is expected to exhibit
mixing induced $CP$ violation.   
This final state can
be accessed by both $B^0$ and $\overline{B^0}$.
$CP$ violation in this case would manifest itself as:
\begin{equation}
 {dN\over{dt}}(B^0\rightarrow J/\psi K^0_S) \ne 
 {dN\over{dt}}(\overline{B^0}\rightarrow J/\psi K^0_S)
\end{equation}
where $J/\psi = |c\overline{c}>$,  
$K^0_S = \frac{1}{\sqrt{2}}(|d\overline{s}>+|s\overline{d}>)$
and the final state, $J/\psi K^0_S$ is a CP eigenstate:
\begin{equation}
CP|J/\psi K^0_S> = -|J/\psi K^0_S >
\end{equation}

\begin{figure}[t]
\begin{center}
\epsfxsize=33pc 
\hspace*{0.2truein}
\epsfbox{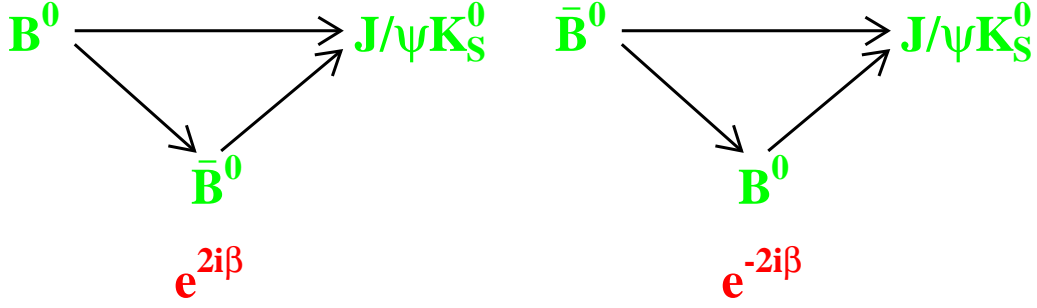}
\caption{$CP$ violation via mixing.  Since both $B^0$ and $\overline{B^0}$
can decay to the $CP$ eigenstate $J/\psi K^0_S$, we have an interference
between the mixed and unmixed decays.  For example, if a meson is produced
as a $B^0$ (shown on the left) and ultimately decays into $J/\psi K^0_S$,
it could have decayed as a $B^0$ or mixed into a $\overline{B^0}$ 
before decaying.  The process shown on the right is for an initial
state $\overline{B^0}$.  The interference exposes the phase in the CKM
matrix element $V_{td}$, giving rise to $CP$ violation in the Standard
Model.}
\label{fig:cpv}
\end{center}
\end{figure}

In the CKM framework, $CP$ violation occurs in this mode
because the mixed decay and direct decay interfere with one
another. This is shown in Fig.~\ref{fig:cpv}.  An initial state
$B^0$ can decay directly to $J/\psi K^0_S$, or it can mix into
a $\overline{B^0}$ and then decay to $J/\psi K^0_S$.  The interference
between those two paths exposes the complex phase in the CKM matrix
element $V_{td}$.

When we  produce $\overline{B^0}$ at $t=0$:
\begin{equation}
\label{eq:b0bar}
{dN\over{dt}}(\overline{B^0}\rightarrow J/\psi K^0_s) \propto
     e^{-t/\tau}(1+\sin2\beta \sin\Delta m_d t) 
\end{equation}
\noindent If we  produce $B^0$ at $t=0$:
\begin{equation}
\label{eq:b0}
{dN\over{dt}}(B^0\rightarrow J/\psi K^0_s) \propto
     e^{-t/\tau}(1-\sin2\beta \sin\Delta m_d t)
\end{equation}
\noindent Forming the asymmetry:
\begin{displaymath}
\label{eq:asym}
   A_{CP}(t)= {{dN\over{dt}}(\overline{B^0} \rightarrow J/\psi K^0_S) -
               {dN\over{dt}}(B^0 \rightarrow J/\psi K^0_S)\over {
               {dN\over{dt}}(\overline{B^0} \rightarrow J/\psi K^0_S) +
               {dN\over{dt}}(B^0 \rightarrow J/\psi K^0_S)}} 
\end{displaymath}

\begin{equation}
 A_{CP}(t) =  \sin2\beta \ \sin(\Delta m_dt).  
\end{equation}
\noindent This is the time-dependent equation for the $CP$ asymmetry
in this mode.   The asymmetry as a function of proper time
oscillates with a frequency of $\Delta m_d$.  The amplitude
of the oscillation is $\sin 2 \beta$, where $\beta$ is the
angle of the unitarity triangle shown earlier.

We can also perform the time-integral of equation~\ref{eq:asym}:

\begin{eqnarray}
   A_{CP}\! \! & =\! \!  
           & \! \!
 {\int {dN\over{dt}}(\overline{B}^0 \rightarrow \psi K^0_S)dt -
                  \int {dN\over{dt}}(B^0 \rightarrow \psi K^0_S)dt \over {
                  \int {dN\over{dt}}(\overline{B}^0 \rightarrow \psi K^0_S)dt +
                  \int {dN\over{dt}}(B^0 \rightarrow \psi K^0_S)}dt} \\
    A_{CP} &  = & {N(\overline{B}^0 \rightarrow \psi K^0_S) -
                   N(B^0 \rightarrow \psi K^0_S)\over {
                   N(\overline{B}^0 \rightarrow \psi K^0_S) +
                   N(B^0 \rightarrow \psi K^0_S)}}   \\
\end{eqnarray}
Integrating equations~\ref{eq:b0} an \ref{eq:b0bar} and substituting
them, we get:
\begin{eqnarray}
   A_{CP} & = & {\Delta m_d \tau_{B}
   \over{1+(\Delta m_d \tau_{B})^2}} \cdot \sin2\beta  \ \\
    A_{CP}       &  \simeq &
           0.47\sin2\beta   
\end{eqnarray}

This shows that we do not need to measure the proper time of
the events.  Integrating over all lifetimes still yields an 
asymmetry, although information is lost in going from the time
dependent to the time-integrated asymmetry.  The above
formalism is true when the $B^0$ and $\overline{B^0}$ are 
produced in an incoherent state, as they are in high energy
hadron collisions.  At the $\Upsilon(4s)$, the $B^0$ and
$\overline{B^0}$ are produced in a coherent state and the
time-integrated asymmetry vanishes.\cite{somebody}

\subsection{Experimental Issues}

The bottom line when it comes to $CP$ violation in the 
$B$ system is that you need to tell the difference between
$B^0$ mesons and $\overline{B^0}$ mesons at the time of
production.   After identifying a sample of signal events,
flavor tagging is the most important aspect of analyses of
$CP$ violation.

In the case of the $J/\psi K^0_S$ final state, we have 
no way of knowing whether the meson was a $B^0$ or 
$\overline{B^0}$ as it decayed, nor do we need to know.
The difference we are attempting to measure is the decay
rate difference for mesons that were \bf produced \rm
as $B^0$ or $\overline{B^0}$.  In this case, we are tagging
the flavor of the $B$ meson when it was produced.

The analysis we are going to discuss here is a measurement of 
the $CP$ asymmetry in $B^0/\overline{B^0}\rightarrow J/\psi K^0_S$
from the CDF experiment.
Before discussing that measurement, we begin with by 
presenting some of the
unique aspects to $b$ 
physics in the hadron collider environment.

\subsubsection{$B$ Production and Reconstruction}

First of all, the $b\overline{b}$ cross section is enormous,
${\cal O}(100 \mu b)$, which means at typical operating luminosities,
1000 $b\overline{b}$ pairs are produced every second!  The $b\overline{b}$
quarks are produced by the strong interaction, which preserves
``bottomness'', therefore they are always produced in pairs.
The transverse momentum ($p_T$) spectrum for the produced $B$ hadrons is 
falling very rapidly, which means that most of the $B$ hadrons 
have very
low transverse momentum.  For the sample of $B\rightarrow J/\psi K^0_S$
decays we are discussing here, the average $p_T$ of the $B$ meson is
about $10\, \rm  GeV/c$.  The fact that the $B$ hadrons have low transverse
momentum does not mean that they have low total momentum.  Quite frequently,
the $B$ mesons have very large longitudinal momentum (longitudinal being
the component along the beam axis.)  These $B$ hadrons are boosted along
the beam axis and are consequently outside the acceptance of the detector.

\begin{figure}[ht]
\begin{center}
\epsfxsize=4truein 
\hspace*{0.5truein}
\epsfbox{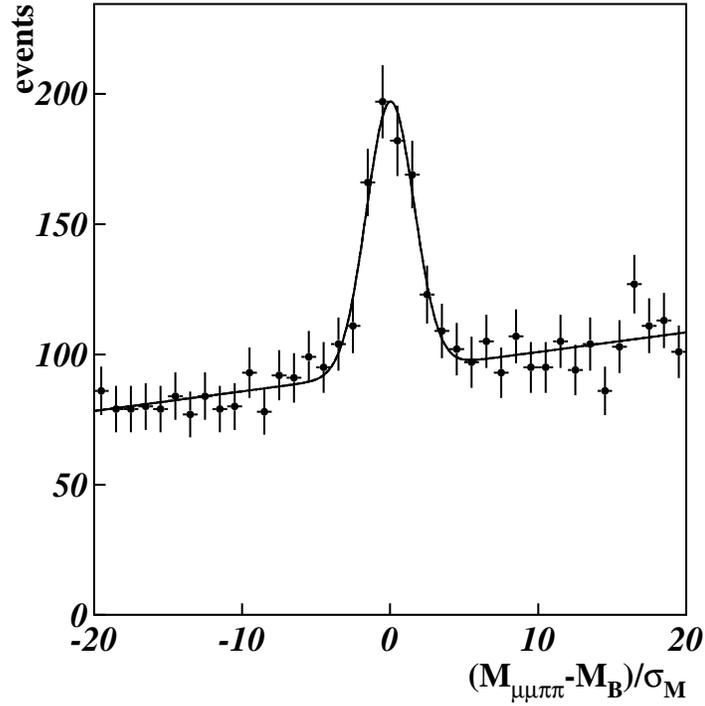}
\caption{$B^0/\overline{B^0}\rightarrow J/\psi K^0_S$ event yield after
the selection criteria discussed in the text have been applied.  The
data is plotted in units of ``normalized mass'': $m_{norm}=(M_{fit}-
M_{B})/\sigma_{fit}$, where $M_{fit}$ and $\sigma_{fit}$ are the 
four track fitted mass and uncertainty, respectively, and $M_B$ is
the world average $B^0$ mass.  Signal events show up with $M_{norm}$
near zero, while combinatoric background shows up uniformly across
the plot.
}
\label{fig:psikshort}
\end{center}
\end{figure}

\setcounter{footnote}{0}

For $b\overline{b}$ production, 
like $W$ production discussed previously, 
the center of mass of the
parton-parton collision is not at rest in the lab frame.
Even in the cases where one $B$ hadron is reconstructed (fully or 
partially) within the detector, the second $B$ hadron may be outside
the detector acceptance.  

To identify the $B$ mesons, we must first trigger the detector readout.
Even though the $b\overline{b}$ production rate is large, it is about
1000 times below the generic inelastic scattering rate.  In the trigger,
we attempt to identify leptons: electrons and muons.  In this analysis,
we look for two muons, indicating that we may have had a $J/\psi\rightarrow
\mu^+ \mu^-$ decay.\footnote{It is  difficult
to trigger on the decay $J/\psi \rightarrow e^+ e^-$ at a hadron
collider.   The distinct  aspect of electrons is their energy 
deposition profile in the calorimeter.  For low $p_T$  electrons 
from $J/\psi$ decays ($p_T<10\, \rm GeV/c$), 
there is sufficient overlap from other particles
to cause high trigger rates and low signal-to-noise.}

Once we have the data on tape, we can attempt to fully reconstruct the
$B^0/\overline{B^0}\rightarrow J/\psi K^0_S$ final state.  The
event topology that we are describing here can be seen in 
Fig.~\ref{fig:tags}.  To reconstruct $B\rightarrow J/\psi K^0_S$,
we again look for $J/\psi \rightarrow \mu^+\mu^-$, this time with 
criteria more stringent than those imposed by the trigger.  Once we 
find a dimuon pair with invariant mass consistent with the $J/\psi$ mass,
we then look for the decay $K^0_S\rightarrow \pi^+\pi^-$.  At this
point, we require the dipion mass be consistent with a $K^0_S$ mass,
and we also take advantage of the fact that the $K^0_S$ lives a 
macroscopic distance in the lab frame.  Once we have both a $J/\psi$
and $K^0_S$ candidate, we put them all together to see if they were
consistent with the decay $B^0/\overline{B^0}\rightarrow J/\psi K^0_S$.
For example, the momentum of the $K^0_S$ must point back to the 
$B$ decay vertex, and the $B$ must point back to the primary (collision)
vertex.  After all of these selection criteria, we have a sample of 
400 signal events with a signal to noise of about 0.7-to-1, as shown
in Fig~\ref{fig:psikshort}.

\subsubsection{Flavor Tagging and Asymmetry Measurement}
 
Now that we have a sample of signal events (intermixed with background),
we must attempt to identify the flavor of the $B^0$ or $\overline{B^0}$ 
at the time of production using the flavor tagging techniques outlined
above.  For this analysis, we use three techniques: same-side tagging,
lepton tagging and jet charge tagging.  The lepton and jet charge flavor
tags are looking at information from the other $B$ hadron in the event
to infer the flavor of the $B$ we reconstructed.  Table~\ref{ta:taggers}
summarizes the flavor tagging efficiency and dilution for each of
the algorithms.

\begin{table}[thb]
\caption{Summary  of tagging algorithms performance.
All numbers listed are in percent. 
The efficiencies  are obtained from the $B \to J/\psi K^0_S$ sample.
The dilution information
is derived from the $B^\pm \to J/\psi K^{\pm}$ sample.
}
\label{ta:taggers}
\begin{center}
\begin{tabular}{|ccccc|}
\hline
tag side &tag type & efficiency ($\epsilon$)& dilution (${\cal D}$)
& $\epsilon {\cal D}^2$  \\
\hline 
same-side &
same-side &
$73.6 \pm 3.8$ & $16.9\pm 2.2$ & $2.1\pm 0.5$ \\
opposite side &
soft lepton & 
$5.6\pm 1.8$ &
$62.5 \pm 14.6$ & $2.2\pm 1.0$ \\
& 
jet charge 
&$40.2 \pm 3.9$ & 
 $23.5 \pm 6.9$ & $ 2.2 \pm 1.3 $  \\

\hline

\end{tabular}
\end{center}
\end{table}

With the sample of events, the proper decay time and the measured
flavor for each event, we are ready to proceed.  
In practice, we are measuring ${\cal A}(t)$: 
\begin{equation}
{\cal A}(t)  = {1\over{{\cal D}}} \left({N_--N_+\over{N_-+N_+}}
\right)
 = {1\over{{\cal D}}} A_{raw}(t)
\end{equation}
where $N_-$($N_+$) are the number of negative (positive) 
tags.  A negative tag indicates a $\overline{B^0}$, while
a positive tag indicates a $B^0$.  We do not write $\overline{B^0}$ and
$B^0$ in the equation, though, because not every negative
tag is truly a $\overline{B^0}$.

We arrive at the quantity ${\cal A}_{raw}$
using the $J/\psi K^0_S$ data, but to get to the measured asymmetry,
we must also know ${\cal D}$.  We can measure ${\cal D}$ 
using control samples and
Monte Carlo, but it can not be extracted from the $J/\psi K^0_S$ data.
Since typical dilutions are about $20\%$, that means that the 
raw asymmetry is 1/5 the size of the measured asymmetry.  The 
higher the dilution (the more effective the flavor tagging method)
the closer the raw  asymmetry is to the measured  asymmetry.
We can classify the statistical uncertainty on
the asymmetry as:
\begin{equation}
\label{eq:da}
(\delta {\cal A})^2 = (\delta {\cal A}_{raw}/{\cal D})^2 + 
  ({\cal A}_{raw}/{\cal D})^2 (\delta {\cal D}/{\cal D})^2     
\end{equation}
where $\delta {\cal D}$ is the uncertainty on the dilution and
$\delta {\cal A}_{raw}$ is the statistical uncertainty
on the raw  asymmetry.  Ignoring (for the moment) the
presence of background in our sample,  
$(\delta {\cal A}_{raw})_{stat} = 
1/\sqrt{N_{tagged}} = 1/\sqrt{\epsilon N_{sig}}$, where $\epsilon$
is the flavor tagging efficiency discussed previously and $N_{sig}$
is the number of signal events.
More realistically, we can not neglect the presence of background,
and the statistical uncertainty on the measured asymmetry is:
$(\delta {\cal A}_{raw})_{stat} 
= {1\over{\sqrt{\epsilon N_{sig}}}}\sqrt{{N_{sig}+B \over{N_{sig}}}}$.
The first term
in Equation \ref{eq:da} is the ``statistical'' uncertainty on
the asymmetry and is of the form: $\delta {\cal A} = 1/
\sqrt{\epsilon {\cal D}^2 N_{sig}}$.  Not only does the dilution 
factor degrade the raw  asymmetry, it also inflates the 
statistical error.  Think of it this way: we have events that
we are putting into two bins--a $B^0$ bin and a $\overline{B^0}$ 
bin.  When we tag an event incorrectly (mistag), we take it out
of one bin and put it into the other bin.  Not only do we have 
one less event in the correct bin, we have one more event in the
incorrect bin!  This hurts our measurement more than had we simply
removed the event from the correct bin and thrown it away.

In reality, there are several complications to this measurement:
\begin{itemize}
 \item Our data sample has both signal and background events in it.
For an  event in the signal region, we don't know \it a priori \rm
if it is signal or background.
 \item We are using multiple flavor tagging algorithms.  Each 
algorithm has a different ${\cal D}$ associated with it.  Some events are
tagged by more than one algorithm, and those two tags may agree or
disagree.
  \item  Due to experimental acceptance, 
 not every event in our sample has a precisely determined proper decay
 time.
 \item Due to experimental acceptance, the efficiencies for positive and
 negative tracks are not identical (although the correction factor is tiny.)
\end{itemize}
We handle  these effects with a maximum likelihood fit that 
accounts for the probability that any given event is signal versus 
background and tagged correctly versus incorrectly.  In doing so,
we not only account for the multiple flavor tagging algorithms and
the background in our data, but the correlations between all of 
these elements is handled as well.\cite{prd}

\begin{figure}[ht]
\begin{center}
\epsfxsize=4truein 
\hspace*{0.5truein}
\epsfbox{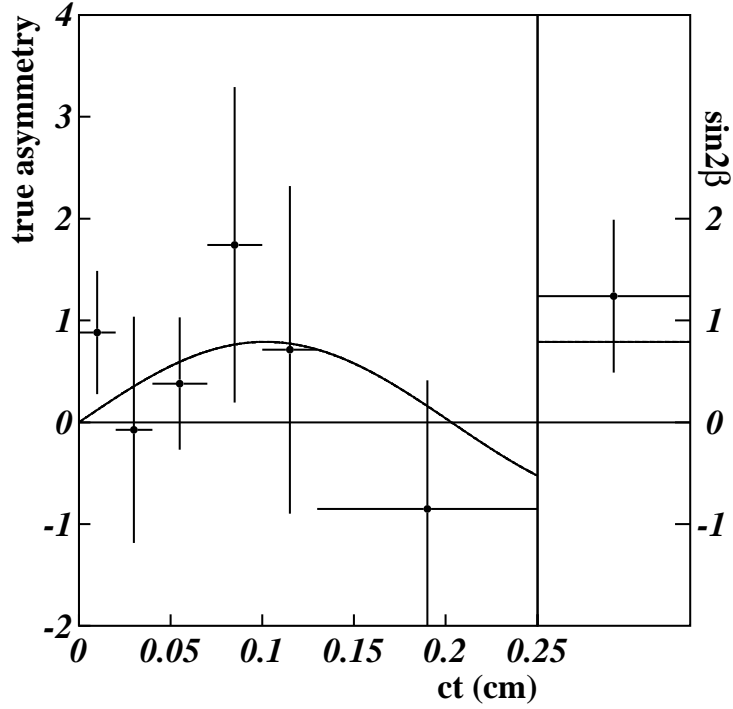}
\caption{The true asymmetry 
(${\cal A}_CP(t) = \sin 2 \beta \sin \Delta m_d t$) as a function of lifetime
for $B\rightarrow J/\psi K^0_S$ events.  The data points
are sideband-subtracted and have been combined according
to the effective dilution for single and double-tags.
The events are shown in the rightmost point are those that
do not have precision lifetime information.}
\label{fig:asym}
\end{center}
\end{figure}

\subsubsection{Results}

The final result of our analysis is show in Fig.~\ref{fig:asym}.
The points 
are the $J/\psi K^0_S$
data, after having subtracted out the contribution from the background.
The data has also been corrected for the flavor tagging dilutions.
The solid curve is the fit to the data of the functional form:
${\cal A}_{CP} = \sin 2\beta \sin\Delta m_d t$, with $\Delta m_d$
constrained to the world average value.  The amplitude of the 
oscillation is $\sin 2 \beta$.  The single point to the right shows
all events that do not have a precisely measured lifetime.  As shown
earlier, the time-integrated asymmetry is nonzero, therefore these
events are quite useful in extracting $\sin 2 \beta$.  

The result of this analysis is:
\begin{displaymath}
\sin2\beta = 0.79 {+0.41 \atop -0.44}\, \rm (stat.+syst.)
\end{displaymath}
This is consistent with the expectation of $\sin 2\beta =0.75$
based upon indirect fits to other data.  This result
rules out $\sin 2\beta = 0$ at the 93\% confidence level, not
sufficient to claim observation of $CP$ violation in the $B$
system.  On the other hand, this is the best direct evidence
to date for $CP$ violation in the $B$ system.  When broken down
into statistical and systematic components, the uncertainty is
$\delta(\sin 2 \beta) = \pm 0.39 (\rm stat.) \pm  0.16 (\rm syst.)$.
The total uncertainty is dominated by the statistics of the sample and 
efficacy of the flavor tagging.  The systematic uncertainty arises
from the uncertainty in the dilution measurements ($\delta {\cal D}$.)
However, the uncertainty on the dilution measurements are actually
limited by the size of the data samples used to measure the dilutions.
In other words, the \bf systematic \rm uncertainty on $\sin 2 \beta$ 
is really a \bf statistical \rm uncertainty on the ${\cal D}$'s.
As more data is accumulated in the future, both the statistical and
systematic uncertainty in $\sin 2 \beta$ will decrease as $1/\sqrt{N}$.

\begin{figure}[t]
\begin{center}
\hspace*{0.4truein}
\epsfxsize=4truein 
\hspace*{0.5truein}
\epsfbox{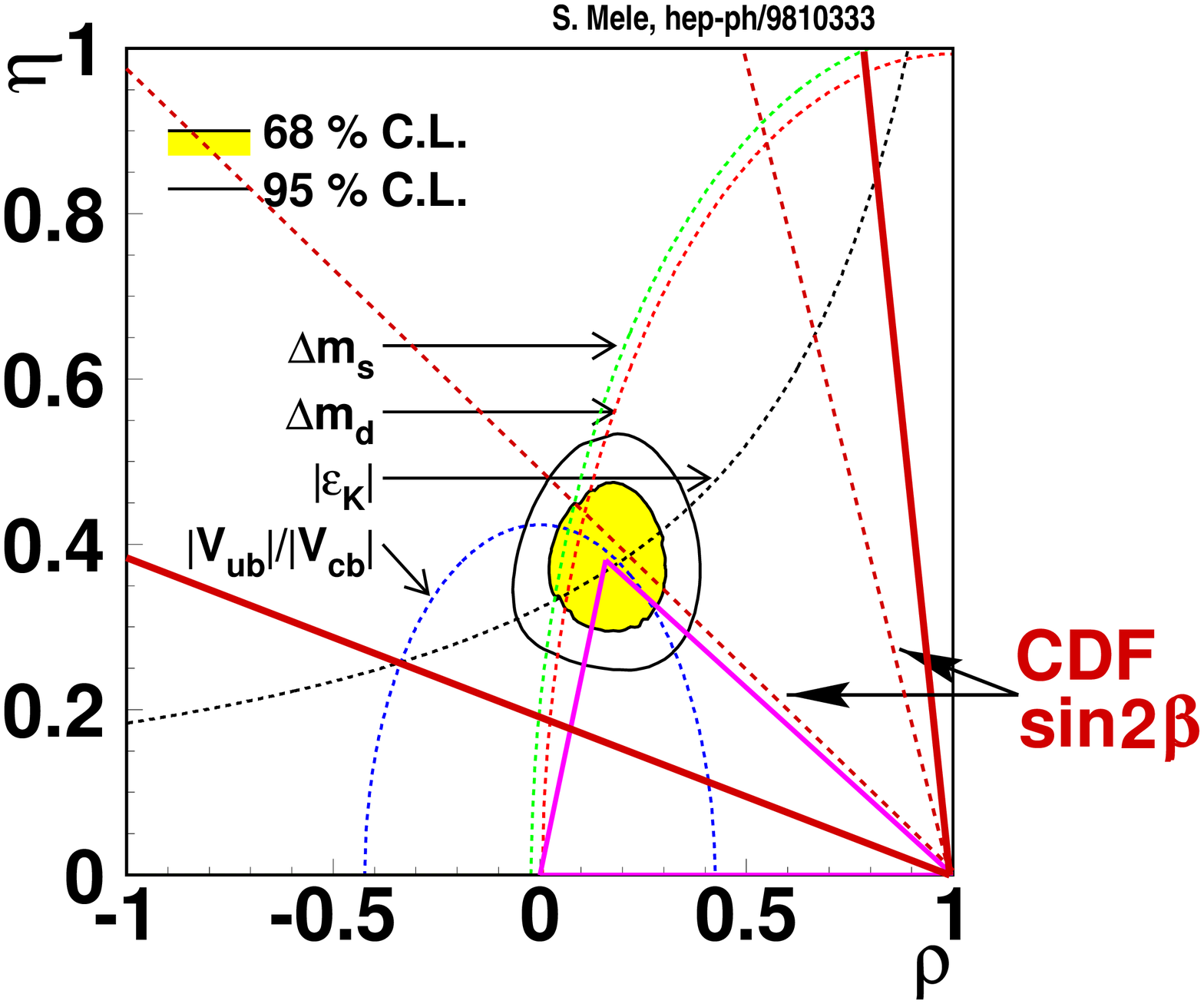}
\caption{The experimental determination of $\rho$ and $\eta$.  The
curves are based upon experimental measurements of $V_{ub}$,
$\epsilon_K$, $B^0_d$ and $B^0_s$ mixing.  
The contours are the result
of the global fit to the data.\cite{mele}
The dashed
lines originating at (1,0) 
are the two solutions for $\beta$ corresponding to
$\sin 2 \beta = 0.79$.  The solid lines are the $1\sigma$ contours
for this result.}
\label{fig:rhoeta}
\end{center}
\end{figure}

Figure~\ref{fig:rhoeta} shows the contours which result from
global fits to measured data in the $B$ and $K$ 
system.\cite{nierste,mele} 
 The dashed
lines originating at (1,0) 
are the two solutions for $\beta$ corresponding to
$\sin 2 \beta = 0.79$.  The solid lines are the $1\sigma$ contours
for this result.  Clearly the result shown here is in good agreement
with expectations.

The uncertainty on the $\sin 2\beta$ result presented here
is comparable to the uncertainties from 
recent measurements by the Belle and Babar 
collaborations\cite{somebody,somebody2}.
While none of the measurements are yet to have the precision
to stringently test the Standard Model, the fact that  this
measurement can be made in two very different ways is interesting.
The hadron collider environment has an enormous $b\overline{b}$
cross section, but backgrounds make flavor tagging difficult.  In
the $e^+e^-$ environment, the production cross section is 
much smaller, but the environment lends itself more favorably
to flavor tagging.   These facts make the measurements performed
in the different environments complementary to one another.

\subsection{The Future}

The Fermilab Tevatron is scheduled to Run again in 2001.
Both CDF\cite{cdf2} and D\O\cite{d02}  
detectors are undergoing massive upgrades
in order to handle more than a factor of 20 increase in data.
In addition, $e^+e^-$ $B$-factories at Cornell (CLEO-III),\cite{cleo} 
KEK (Belle)\cite{belle} and SLAC (BaBar)\cite{babar}
are all currently taking data.  Finally, Hera-B,\cite{herab}  a dedicated
$B$ experiment at DESY, also will begin taking data in 2001.

On the timescale of 2003-2004, there could be as many as 5 
different measurements of $\sin 2 \beta$, all of them with
an uncertainty of  $\delta (\sin 2 \beta) \ltsim 0.1$.  
Putting these together would yield a world average measurement
with an uncertainty of  $\delta (\sin 2 \beta) \ltsim 0.05$.
Although this alone will provide an impressive constraint upon
the unitarity triangle, it will not be sufficient to 
thoroughly  test the Standard Model for self-consistency.
On the same timescale, improvements are required in the lengths
of the sides
of the triangle, as well as other measurements of the angles.
Finally, there are measurements of other quantities that are
not easily related to the unitarity triangle that are important
tests of the Standard Model.

The following is a list of some of the measurements
that will be undertaken and/or improved-upon in the coming years
 (\it this is an incomplete list\rm ):
\begin{itemize}
\item $CP$ asymmetries in other modes:
\it e.g. \rm
\begin{itemize}
\item $B^0/\overline{B^0}\rightarrow \pi^+\pi^-$;
\item $B^0_s/\overline{B^0_s}\rightarrow J/\psi \phi$; 
\item  $B^0_s/\overline{B^0_s}\rightarrow K^+ K^- $; 
\item  $B^0_s/\overline{B^0_s}\rightarrow D_s^\pm K^\mp $; 
\item  $B^0/\overline{B^0}\rightarrow D^+D^-$.
\end{itemize}
\item $B^0_s/\overline{B^0_s}$ mixing.
\item rare $B$ decays: \it e.g. \rm 
$B^\pm \rightarrow \mu^+\mu^- K^\pm$; $B^0\rightarrow
\mu^+\mu^-$.
\item radiative $B$ decays: \it e.g. \rm 
$B^0 \rightarrow K^*\gamma$; $B^0_s \rightarrow \phi \gamma$.
\item improved measurements of $V_{ub}$: \it e.g. \rm $B\rightarrow \pi \pi$;
$B\rightarrow \rho \ell \nu$.
\item mass and lifetime of the $B_c$ meson.
\item mass and lifetimes of the $B$ baryons: \it e.g. \rm $\Lambda_b = |udb>$.
\end{itemize} 

It will take many years and a body of measurements to gain
further insights into the mechanisms behind the CKM matrix
and $CP$ violation.

Advances in kaon physics over the last 40 years and advances
in $B$ physics in the last 25 years have put us on track
to carry out these measurements in the very near future.
These measurements will hopefully bring us
to a more fundamental understanding to the mechanism behind
$CP$ violation.

\section*{Acknowledgments}

I would like to thank David Burke, Lance Dixon, 
Charles Prescott 
and the organizers
of the 2000 SLAC Summer Institute.
I would also like to thank and 
acknowledge the collaborators of the CDF and D\O\ experiments,
as well as the Fermilab accelerator division.
This work is supported by the U.S. Department
of Energy Grant Number DE-FG02-91ER40677.








%


\end{document}